\documentclass[preprint,authoryear,12pt]{elsarticle}

\usepackage{amssymb}
\usepackage{url}
\usepackage{graphicx}
\usepackage{amsmath}
\usepackage[margin=2.3cm]{geometry}
\usepackage{threeparttable}

\usepackage{url}
\usepackage{lineno} 
\usepackage{color}
\usepackage{xcolor}
\usepackage{soul}
\usepackage{caption}
\usepackage{subcaption}
\usepackage{lscape}
\usepackage{comment} 
\usepackage{colortbl} 
\usepackage{listings}
\usepackage{hyperref}
\usepackage{blkarray} 
\usepackage{bbm}
\usepackage{bm}
\modulolinenumbers[5]

\usepackage{array} 

\usepackage{setspace, verbatim, mathrsfs}
\usepackage{MnSymbol}
\usepackage{dsfont, bm}
\usepackage{ulem}
\usepackage{cancel}

\usepackage{mymacros} 

\usepackage{empheq} 

\usepackage{tcolorbox}

\newcolumntype{L}[1]{>{\raggedright\let\newline\\\arraybackslash\hspace{0pt}}m{#1}}
\newcolumntype{C}[1]{>{\centering\let\newline\\\arraybackslash\hspace{0pt}}m{#1}}
\newcolumntype{R}[1]{>{\raggedleft\let\newline\\\arraybackslash\hspace{0pt}}m{#1}}

\definecolor{ao}{rgb}{0.0, 0.5, 0.0}
\definecolor{bred}{rgb}{0.8, 0.0, 0.0}
\definecolor{carrotorange}{rgb}{0.93, 0.57, 0.13}

\makeatletter
\def\ps@pprintTitle{%
  \let\@oddhead\@empty
  \let\@evenhead\@empty
  \let\@oddfoot\@empty
  \let\@evenfoot\@oddfoot
}
\makeatother

\journal{Computational Statistics \& Data Analysis}

\definecolor{jay}{rgb}{0.39, 0.05, 0.05}

\begin{document}

\begin{frontmatter}

\title{Bayesian spatio-temporal models for stream networks}

\author[qut,acems, myfootnote]{Edgar Santos-Fernandez \corref{cor1}}
\ead{edgar.santosfdez@gmail.com}
\corref{edgar.santosfdez@gmail.com}
\fntext[myfootnote]{School of Mathematical Sciences. Y Block, Floor 8, Gardens Point Campus. Queensland University of Technology
GPO Box 2434. Brisbane, QLD 4001. Australia}
\author[noaa]{Jay M. Ver Hoef} 
\author[erinp,qut,acems]{Erin E. Peterson}
\author[qut]{James McGree}
\author[us]{Daniel J. Isaak}
\author[qut,acems]{Kerrie Mengersen}


\cortext[cor1]{Corresponding author}

\address[qut]{School of Mathematical Sciences. Queensland University of Technology}
\address[acems]{Australian Research Council Centre of Excellence for Mathematical and Statistical Frontiers (ACEMS)}
\address[noaa]{Marine Mammal Laboratory. NOAA-NMFS Alaska Fisheries Science Center}
\address[erinp]{Erin Peterson Consulting. Brisbane Australia}
\address[us]{Rocky Mountain Research Station. US Forest Service}

\begin{abstract}
Spatio-temporal models are widely used in many research areas including ecology. The recent proliferation of the use of \textit{in-situ} sensors in streams and rivers supports space-time water quality modelling and monitoring in near real-time. 
A new family of spatio-temporal models is introduced. These models incorporate spatial dependence using stream distance while temporal autocorrelation is captured using vector autoregression approaches.
Several variations of these novel models are proposed using a Bayesian framework.
The results show that our proposed models perform well using spatio-temporal data collected from real stream networks, particularly in terms of out-of-sample RMSPE. This is illustrated considering a case study of water temperature data in the northwestern United States.

\end{abstract}

\begin{keyword}
Bayesian model \sep space-time \sep linear regression \sep branching network \sep vector autoregression
\end{keyword}

\end{frontmatter}

\section {Introduction}
\label{sec:Int}
Freshwater ecosystems represent only 2.3\% of the Earth’s surface, but host 9.5\% of the described animal species on Earth \citep{reid2019emerging}. These biodiversity hotspots are arguably the most endangered ecosystems on Earth, with rapid population declines due to habitat degradation, water pollution, and modification of water flows, among other factors \citep{dudgeon2006freshwater, reid2019emerging}. A lack of data has hindered science-based management in the past, but the advent of \textit{in-situ} sensing in streams and rivers is increasing the density and volume of environmental monitoring data. 
For example, it is now possible to semi-continuously (e.g. every 15 minutes) monitor water quality \citep[e.g.,][]{ficklin_effects_2013,stackpoole_spatial_2017}, nutrient dynamics \citep{wollheim2017aquatic} and
fisheries \citep[e.g.,][]{isaak2017scalable}.
High-frequency data from multiple spatial locations are likely to exhibit spatio-temporal dependencies that reflect the topology of the stream network (e.g. branching network structure, connectivity, directional water flow and volume), but few methods exist that describe this unique spatio-temporal autocorrelation. Thus, our aim is to develop Bayesian spatio-temporal models for data on stream networks that accommodate their unique spatio-temporal characteristics and then use a real dataset to demonstrate how these models provide additional probabilistic information for the management of a threatened freshwater species.

Statistical models for spatio-temporal datasets collected on streams range from pure time-series models, spatial analyses, and a combination of both. In each of these cases, the unique branching structure of the stream network, and the directionality and volume of water flowing from upstream to downstream are accounted for to different degrees.  For example, several standard time-series models have been developed to capture temporal autocorrelation in the error term including a first-order autoregressive model AR(1)  \citep[e.g.,][]{bal2014hierarchical, hague2014evaluation, letcher2016hierarchical} and autoregressive integrated moving average (ARIMA) models \citep[e.g.][]{graf2018distribution}. The central purposes of these applications were typically temporal interpolation and forecasting future values at sensor locations thus, there was no need to describe unique spatial relationships on streams.

In contrast, there have been numerous advances in spatial regression methods for stream networks that describe spatial dependence between locations based on in-stream distances (i.e. distance travelled along the stream network), directionality, and flow volume and also allow for spatial prediction (i.e. kriging) throughout the branching network. \citet{ver2006spatial} introduced several covariance structures built using spatial moving averages and  \citet{cressie2006spatial} also suggested methods to account for spatial autocorrelation.   
\citet{ver2006spatial} methods have been used across a variety of applications  \citep[e.g.,][]{isaak2014applications, mcmanus2020variation, rodriguez2019spatial} for both estimating regression effects and prediction in locations where data were not observed. In addition, a broad framework for spatial autocovariances that considers branching stream networks and flow connectivity was proposed by \citet{ver2010moving}.

While numerous methods have been used to describe temporal or spatial dependency in streams data, only a small number of spatio-temporal models have been developed for stream network data. \citet{money2009using} developed models that are similar to  \citet{ver2006spatial} using a Bayesian Maximum Entropy (BME) approach to predict dissolved oxygen in the water. Spatial dependence was described using both Euclidean and in-stream distance to compute separable and isotropic, space-time covariance matrices based on an exponential structure for space and exponential/cosinusoidal structure for time. In another example, \citet{money2009modern} used a similar space-time model to estimate \textit{Escherichia Coli (E. coli)} concentrations using BME, based on turbidity measurements at unmeasured locations.  
A spline-based approach to spatio-temporal modelling was also developed by \citet{o2014flexible} and then extended by \citet{jackson2018spatio}, whereby temporal dependence was incorporated in the error term using an AR(1) process. More recently \citet{tang2020space} developed a new family of (I) non-separable covariance structures using space embedding approaches and (II) separable models constructed from valid space and time covariance functions.

Most of the time series approaches used in stream networks are univariate in nature assuming independence between sites, which is a strong assumption. 
Several of the space-time models previously proposed use a static or descriptive representation employing space-time covariance matrices. 
Statistical methods that consider the evolution through time of measurements in stream networks remain largely unexplored.

The general spatio-temporal literature delves into the full separable space-time covariance matrix approach and vector autoregression spatial models. We study both approaches in stream networks settings resulting in separable spatio-temporal stream networks, and vector autoregression spatial stream networks.
We show these two models are equivalent mathematically, but their differing constructions lead to different estimation methods, which we compare using a simple simulated example.

While most of these stream network models embrace a frequentist philosophy, Bayesian alternatives have received far less attention. We exploit the benefits of the Bayesian philosophy e.g.: to produce probabilistic estimates such as exceedance probability, make predictions considering uncertainty, model complex stream network problems, treat more effectively missing data and perform imputation.

Accordingly, our objectives are to: 
\begin{enumerate}
    \item Develop a new family of spatio-temporal models for branching networks, where spatial autocorrelation is incorporated using valid covariance matrices based on in-stream distance along with Euclidean distance approaches. 
    These models will borrow strength across time by employing vector autoregressive models or constructing the full space-time covariance matrices. 
    Our models can be described as Vector Space-Time Stream Network Models (VSTSN). 

    \item Design efficient methods for prediction throughout the branching network and to interpolate/impute missing data in space and time. 
     \item Use Bayesian inference to incorporate uncertainty in the model and to generate probabilistic estimates and predictions and exceedance probabilities. 
    \item Generate new scientific insights in stream temperature regimes by assessing the relationship between temporal dependence and spatial characteristics.  

\end{enumerate}

\section {Methods}
\label{sec:meth}

We start this section with a description of spatial models in the general and broader context and then specifically in stream networks. We then introduce spatio-temporal model variations and describe their formulation using Bayesian hierarchical models. We finalize the section with some performance measures we will use for model selection.   

\subsection{Spatial statistical models based on Euclidean distance}

Linear regression models that describe spatio-temporal dependence are generally formulated as:

\begin{equation}
\pmb{y} = \pmb{X}\pmb{\beta} + \pmb{v} + \pmb{\epsilon},
\label{eq:lm0}
\end{equation}

\noindent where $\pmb{y} = [\pmb{y}_{1}^{'}, \pmb{y}_{2}^{'}, \cdots,\pmb{y}_{T}^{'}]^{'}$
is a stacked response vector of length $n = S\times T$ for $S$ spatial locations and $T$ time points,
$\pmb{X} = [\pmb{X}_1^{'}, \pmb{X}_2^{'},\cdots,\pmb{X}_T^{'}]^{'}$ is a $n \times p$ design matrix of $p$ covariates, $\pmb{\beta}$ is a $p \times 1$ vector of regression coefficients, $\pmb{v}=[\pmb{v}_1^{'}, \pmb{v}_2^{'},\cdots,\pmb{v}_T^{'}]^{'}$ is a vector spatio-temporal autocorrelated random effects, and $\pmb{\epsilon}$ is the unstructured error term; i.e., $\textrm{var}(\pmb{\epsilon}) = \sigma^2_0\pmb{I}$ where $\sigma_{0}^2$ is called the nugget effect and $\pmb{I}$ is the identity matrix.

In the purely spatial case when $T=1$, then $\pmb{v}$  becomes a spatially-autocorrelated random effect, which can be modelled using a Gaussian process or other spatial structures \citep{banerjee2014hierarchical}, and $\pmb{\epsilon}$ is a vector of spatially-indexed independent random-errors.

Covariance models based on Euclidean distance are traditionally used in geostatistics and some of the most common formulations are the exponential, Gaussian, and spherical functions  \citep{cressie2015statistics, banerjee2014hierarchical}: 

\begin{equation}
    \textrm{exponential model,} \ \  C_{ED}(d\mid \pmb{\theta}) = \sigma_e^2 e^{-3d/\alpha_e}, \alpha_e \in (0,\infty), \sigma_e ^ 2 > 0,
    \label{eq:ced}
\end{equation}
\begin{equation}
    \textrm{Gaussian model,} \ \ C_{ED}(d\mid \pmb{\theta}) = \sigma_e^2 e^{-3(d/\alpha_e)^2},
\label{eq:ced2}
\end{equation}
and
\begin{equation}
\textrm{spherical model,} \ \ C_{ED}(d\mid \pmb{\theta}) = \sigma_e^2\left(1-\frac{3d}{2\alpha_e} + \frac{d^3}{2\alpha^3}\right)\mathbbm{1}(d/\alpha \leqslant 1),
\label{eq:ced3}
\end{equation}

\noindent where $d$ is the Euclidean distance between two locations $s_i$ and $s_j$, $\sigma_e^2$ is the partial sill and $\alpha_e$ is the spatial range parameter. 
We refer to the partial sill as the resulting variance after accounting for the nugget effect (i.e. sill minus nugget effect). 
The spatial range describes how fast the covariance decays with distance.

\subsection{Spatial stream network models}
Stream networks have a unique spatial structure characterised by a branching network topology, directional water flow, and differences in flow volume throughout the network \citep[e.g.][]{peterson2013modelling}.

\begin{figure}[htbp]
  \centering
   \includegraphics[width=3.5in]{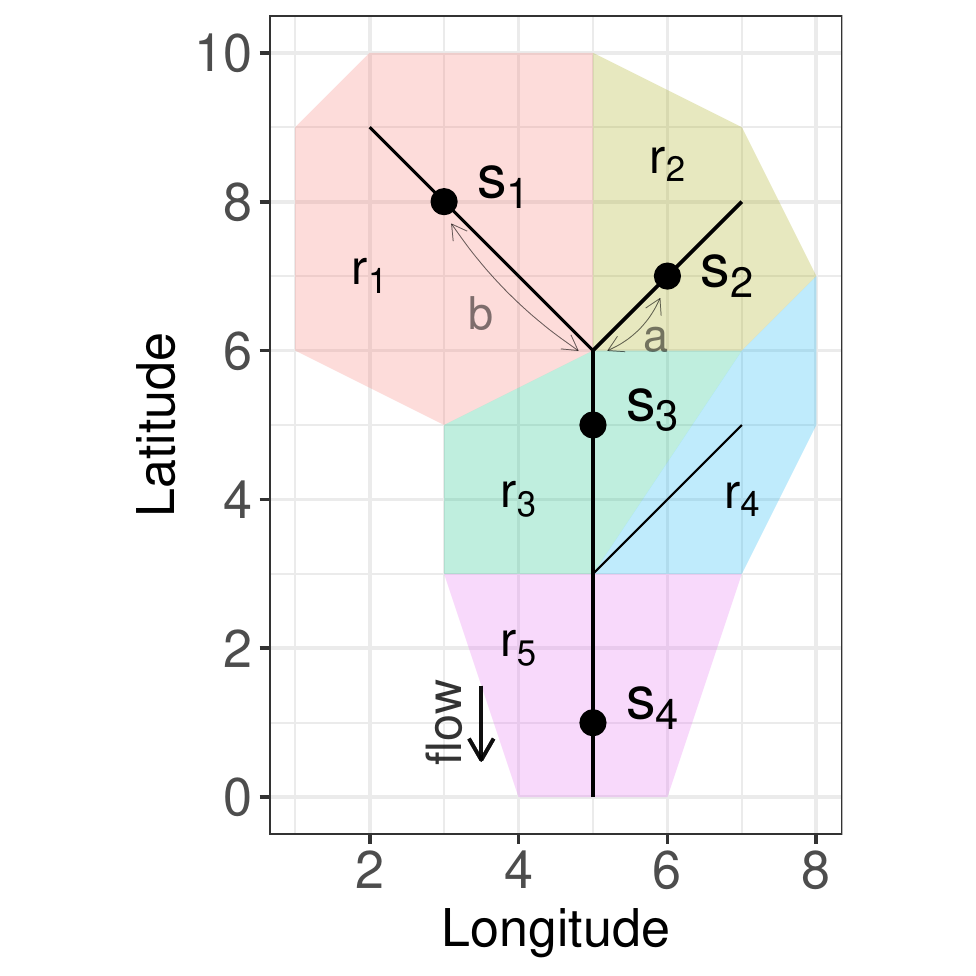}
  \caption{Stream network with four spatial locations ($s_1-s_4$) and five regions ($r_1-r_5$). }
  \label{fig:network}
\end{figure}

Fig.\ref{fig:network} depicts a stream network that we will use for illustration. 
In this first part, let us assume that this is a pure spatial process and concentrate on time $t=1$.
It is composed of five stream segments with five segment contributing areas ($r_1, r_1, \cdots, r_5$) represented via different colors and $S = 4$ spatial locations.
Let $\pmb{y}$ be a vector of random variables at $s = 1, 2, \ldots, S$ unique and fixed spatial locations. 
Here water flows from the top to the bottom. Therefore, spatial locations $s_1$, $s_3$ and $s_4$ are flow-connected, while locations $s_1$ and $s_2$ are flow-unconnected.
The distances from the $s_1$ and $s_2$ to the common confluence are $b$ and $a$ (represented in gray) respectively.

Covariance models based on Euclidean distance, \eqref{eq:ced} - \eqref{eq:ced3}, may not adequately capture spatial dependence in data collected on streams. For example, a hydrologic distance measure (i.e. distance measured \textit{along} the stream network) may more accurately represent proximity in a sinuous stream network than Euclidean distance.

In addition, two locations upstream from a stream junction (i.e. confluence; $s_1$ and $s_2$ in Fig.\ref{fig:network}) may have significantly different water quality because they do not share water flow, even though they reside on the same network and are close in Euclidean space. In contrast, pollutants often move passively downstream and so water quality may be more similar when two locations are flow-connected (e.g. $s_1$ and $s_3$ in Fig.\ref{fig:network}), which occurs when water flows from an upstream location to a downstream location. To address these issues, \citet{ver2010moving} proposed two families of models based on a moving-average construction and found analytical derivations for some autocovariance functions that capture the unique spatial relationships found in streams: tail-up and tail-down models.

\subsubsection{Tail-up models}
In a tail-up model, random variables are constructed by integrating a kernel over a white noise process strictly upstream of a site (i.e. tail of the moving-average function points upstream), which restricts autocorrelation to flow-connected sites. For each of the pairs of sites $s_i$, $s_j$, given that $h$ is the flow-connected hydrologic distance between them, the tail-up covariance matrix can be expressed as

$$C_{TU}(s_i,s_j|\pmb{\theta}) = \left\{\begin{matrix}
0 \:\:\: \textrm{if $s_i,s_j$ are flow-unconnected}, \\C_u(h) W_{ij} \:\:\: \textrm{if $s_i,s_j$ are flow-connected}, \end{matrix}\right.$$ 
where $C_u(h)$ is an unweighted tail-up covariance between two locations, 
and  the $W_{ij}$ represents the spatial weights between sites $i$ and $j$ and is defined by the branching structure of the network. Here, $\pmb{\theta}$ represents the spatial parameters ($\alpha_u$, $\sigma_u$). \citet{ver2010moving} defined a number of unweighted tail-up covariance functions $C_u(h)$:
\begin{align}
\textrm{Tail-up exponential model},& \ \ C_{u}(h \mid \pmb{\theta}) = \sigma_u^2 e^{-3h/\alpha_u}, \\
\textrm{Tail-up linear-with-sill model},& \ \ C_u(h \mid \pmb{\theta}) = \sigma_u^2 (1-h/\alpha_u)\mathbbm{1}(h/\alpha_u \leqslant 1), \\
\textrm{Tail-up spherical model},& \ \ C_u(h \mid \pmb{\theta}) = \sigma_u^2 \left(1-\frac{3h}{2\alpha_u} + \frac{h^3}{2\alpha_u^3}\right)\mathbbm{1}(h/\alpha_u \leqslant 1),
\end{align}
where  $\sigma_u^2$ is the partial sill and $\alpha_u$ is the range parameter. 

An interesting feature of the tail-up model is that to maintain stationary variances we need to split the moving average functions upstream of confluences using spatial weights, $W_{ij}$. Any spatial or ecologically relevant variable can be used to construct the weights, as long as it is available for every segment in the stream network (Fig. \ref{fig:network}). Simple options are to use equal weights or Shreve's stream order \citep{shreve1967infinite}, but the watershed area (i.e. area of land that drains downhill to a common point on the stream network) is commonly used as a surrogate for water volume \citep{frieden2014improving}. 
This variable allows generating additive function values ($AFV$) for each segment in the network, which are then used to construct the spatial weights matrix.

In \ref{sec:wei}, we illustrate the computation of the spatial weights matrix using the example from Fig.\ref{fig:network}.

\subsubsection{Tail-down models}
For tail-down models, the moving average function points in the downstream direction and so spatial dependence may occur between both flow-connected and flow-unconnected locations; although the strength of autocorrelation will differ for pairs of flow-connected and flow-unconnected sites that are an equal distance apart \citep{ver2010moving}. While it is possible to use spatial weights in a tail-down model \citep{hoef2014ssn}, it is not strictly necessary given the dendritic nature of stream networks. 

Again let $h$ be flow-connected hydrologic distance, but for flow-unconnected sites, let $a$ and $b$ be the hydrologic distance from each location to their common confluence, and let $a \leqslant b$ (e.g. $s_1$ and $s_2$ in Fig.\ref{fig:network}). The tail-down covariance between locations can then be defined as follows: 
tail-down exponential model,
$$C_{TD}(a,b, h|\pmb{\theta}) = \left\{\begin{matrix}
\sigma_d^2 e^{-3h/\alpha_d} \:\:\: \textrm{if flow-connected,} \\
\sigma_d^2 e^{-3(a+b)/\alpha_d} \:\:\: \textrm{if flow-unconnected,} \end{matrix}\right. $$
tail-down linear-with-sill model,
$$C_{TD}(a,b, h|\pmb{\theta}) = \left\{\begin{matrix}
\sigma_d^2 (1-\frac{h}{\alpha_d})\mathbbm{1}(\frac{h}{\alpha_d} \leqslant 1) \:\:\: \textrm{if flow-connected,} \\
\sigma_d^2 (1-\frac{b}{\alpha_d})\mathbbm{1}(\frac{b}{\alpha_d} \leqslant 1) \:\:\: \textrm{if flow-unconnected,} \end{matrix}\right. $$
and tail-down spherical model,
$$C_{TD}(a,b, h|\pmb{\theta}) = \left\{\begin{matrix}
\sigma_d^2 (1-\frac{3h}{2\alpha_d} + \frac{h^3}{2\alpha_d^3})\mathbbm{1}(\frac{h}{\alpha_d} \leqslant 1) \:\:\: \textrm{if flow-connected,}\\
\sigma_d^2 (1-\frac{3a}{2\alpha_d} + \frac{b}{2\alpha_d})(1-\frac{b}{\alpha_d})\mathbbm{1}(\frac{b}{\alpha_d} \leqslant 1) \:\:\: \textrm{if flow-unconnected,} \end{matrix}\right. $$
where $\sigma_d^2$ is the partial sill, $\alpha_d$ is the range parameter, and $\mathbbm{1}(\cdot)$ is the indicator function, equal to 1 if its argument is true, otherwise it is zero.

\subsection{Covariance mixture approach}
Data collected on stream networks often exhibit multiple patterns of spatial dependency due to climatic gradients, active movement of organisms within and sometimes between networks, within-stream processes, and passive movement of materials, nutrients, and organisms downstream \citep{peterson2013modelling}. 
Recall that in Eq~\ref{eq:lm0}, for the purely spatial case $\pmb{v}$ is a vector of dimension $s$ corresponding to the spatial locations, whose covariance matrix $\pmb{\Sigma} = COV(\pmb{v})$.
A covariance mixture approach is often used to describe complex spatial patterns in $\pmb{v}$ using a Euclidean distance ($e$) component, along with tail-up ($u$) and tail-down ($d$) components, having the following general form:

\begin{equation}
\pmb{\Sigma} = COV(\pmb{v}) = \pmb{C}_{ED} + \pmb{C}_{TU} + \pmb{C}_{TD}   
= \sigma_e^2\pmb{R}_{e}(\alpha_e)+\sigma_u^2\pmb{R}_{u}(\alpha_u) + \sigma_d^2\pmb{R}_{d}(\alpha_d), 
\label{eq:covs}
\end{equation}

\noindent where $\pmb{C}_{ED}$, $\pmb{C}_{TU}$ and  $\pmb{C}_{TD}$ are matrices derived from covariance functions $C_{ED}(d\mid \pmb{\theta})$, $C_{TU}(s_i,s_j\mid \pmb{\theta})$, and $C_{TD}(a,b,h\mid \pmb{\theta})$, respectively, $\sigma^2_{e}$, $\sigma^2_{u}$, and $\sigma^2_{d}$ are the partial sills for Euclidean, tail-up and tail-down functions, respectively. The correlation matrices $\pmb{R}_{u}(\alpha_u)$, $\pmb{R}_{d}(\alpha_d)$ and $\pmb{R}_{e}(\alpha_e)$ are a function of the range parameters $\alpha_u$, $\alpha_d$ and $\alpha_e$ \citep{hoef2014ssn}. 
The total variance in this formulation is equal to the sum of each of the models' partial sills plus the nugget effect.

\subsection{Spatio-temporal models}
\label{stmodel_general}

Two main approaches can be found in the literature to model spatio-temporal autocorrelation: (1) full covariance or descriptive models in which the space-time covariance function is constructed and (2) dynamical model that involves the evolution of a spatial process \citep{wikle2015modern}.    
From the computational point of view, the bottleneck in these methods is the result of having to invert a large covariance matrix \citep{zhang2018practical}.

The first method involves the construction of a full covariance matrix between all the spatial locations at all time points. This covariance is separable if it is equal to the (Kronecker) product of the spatial and the temporal covariance matrices \citep{porcu201930}. 
Making predictions or kriging requires inverting a covariance matrix.
We use the property that the inverse of the full spatio-temporal covariance matrix is equal to the Kronecker product of the inverse of the spatial and the temporal covariance matrices \citep{wikle2019spatio, porcu201930}.
Despite separable models assume no interaction between space and time, they allow substantial computational gains \citep{mitchell2005testing}, 
See, for example, \citet{flaxman2015fast}, who implemented a fast separable space-time model in the probabilistic programming language Stan \citep{carpenter2017stan}.
Additionally, the inverse of some common temporal covariance matrices e.g. autoregressive AR(p) and some vector autoregressive processes variations VAR(p) can be obtained analytically, resulting in a band space matrix, see \ref{sec:invers} for further details.

Dynamical models, also known as time series spatial processes, incorporate first-order Markovian dependence comprising discrete time intervals. They assume that the spatial correlation does not change across time \citep{posa_simple_1993}.

Both approaches have intrinsic strengths but also limitations.
Some drawbacks are associated with the curse of dimensionality and over-parameterization as we discuss later. 
We performed a simulation study using Bayesian inference in \ref{sec:sim} to:

\begin{enumerate}
    \item compare them in terms of prediction accuracy and computational efficiency
    \item explore the options for producing prediction and assess the model interpolation/imputation 
\end{enumerate}

We found the vector autoregression spatial approach to be computationally more efficient since involves the operations with the spatial rather than with the joint space-time covariance matrix. Therefore, we carry our methods using the vector autoregression spatial approach but keeping in mind that the methods here developed can be formulated using full separable covariance matrices.

\subsection{Spatio-temporal stream network models}
\label{stmodel}

In this section, we examine the spatial stream networks at discrete time points.
We consider repeated measures at times $t = 1, 2, \ldots, T$ of the network shown in Fig.\ref{fig:network}.
Here, let $\pmb{y}_t$ be an $S \times 1$ vector of random variables at unique and fixed spatial locations of $s = 1, 2, \ldots S$. 
For continuous response variables we define conditionally a spatio-temporal model as follows:

\begin{equation}
[\pmb{y}_1,\pmb{y}_2,\cdots,\pmb{y}_T] =  \prod_{t=2}^{T}[\pmb{y}_t \mid  \pmb{y}_{t-1}, \pmb{\theta},\pmb{X}_{t},\pmb{X}_{t-1},\pmb{\beta}, \pmb{\Phi}_1, \pmb{\Sigma}][\pmb{y}_1], 
\label{eq:VAR1def}
\end{equation}
where $\pmb{y}_1$ is the process at $t=1$, and
\begin{equation}
[\pmb{y}_t \mid  \pmb{y}_{t-1},\pmb{\theta},\pmb{X}_{t},\pmb{X}_{t-1},\pmb{\beta}, \pmb{\Phi}_1, \pmb{\Sigma}  ] = \mathcal{N}(\pmb{\mu}_{t},\pmb{\Sigma} + \sigma^2_0\pmb{I}),  
\end{equation}
and
\begin{equation}
\pmb{\mu}_{t} = \pmb{X}_{t}\pmb{\beta} + \pmb{\Phi}_1 (\pmb{y}_{t-1} - \pmb{X}_{t-1}\pmb{\beta}),
\label{eq:err_}
\end{equation}
where $\mathcal{N}(·,·)$ is the probability density function of the normal distribution,
$\pmb{\Sigma} = COV(\pmb{v}_i); \forall \ i = 2,\ldots,T$, is the $S \times S$ spatial covariance matrix defined in Eq~\eqref{eq:covs}, and $\pmb{\Phi}_1$ is a $S \times S$ square transition matrix with elements $\phi_{ij}$  that will determine the amount of temporal autocorrelation. For this process to be stable, the maximum modulus of the eigenvalues of $\pmb{\Phi}_1$ must be less than one (see \citet{lutkepohl2005new, tsay2013multivariate, wikle2019spatio} for additional details).  When the number of spatial locations $S$ is large, the number of parameters to estimate in the matrix $\pmb{\Phi}$ becomes prohibitive, and several approaches have been suggested (e.g., restricting correlation to the four nearest neighbors
\citep{wikle1998hierarchical}).   

Equations  \eqref{eq:VAR1def} - \eqref{eq:err_}  form a vector autoregressive model of order one (VAR(1)) with temporal dependence incorporated in the error term via \eqref{eq:err_}. Multiple variations of the VAR general spatial model have been proposed in the traditional spatial domain \citep[e.g.][]{banerjee2014hierarchical,beenstock2019spatial, wikle1998hierarchical, tagle2019non, wikle2019spatio}. \citet{tagle2019non}, for example, considered a VAR(2) to model wind speeds.

\subsection {Vector autoregression spatial model variations}
\label{sec:dst}

In this section, we propose several model variations with varying complexity levels.

{\bf Case 1 (AR)}

In the simplest case, the diagonal elements of $\pmb{\Phi}$ are all equal to $\phi$ and all the off-diagonal ones are set to zero,
\begin{equation}
\pmb{\Phi}_1 = \begin{bmatrix}
\phi & 0 & 0 & \cdots & 0 \\ 
0 & \phi &  0  & \cdots &0 \\ 
0 & 0 &  \phi &\cdots &0 \\ 
\vdots & \vdots &\vdots & \ddots   &\vdots & \\
0 & 0  & 0 & \cdots & \phi \\ 
\end{bmatrix},    \phi \in (-1, 1).
\end{equation}
This simple model assumes the same temporal autocorrelation for all spatial locations, which can be restrictive.

{\bf Case 2 (VAR)}

A second approach allows $\phi$ to be site specific ($\phi_{1}, \phi_{2}, \cdots, \phi_{S}$), which is known as the autoregressive shock model \citep{wikle1998hierarchical},

\begin{equation}
\pmb{\Phi}_1 = \begin{bmatrix}
\phi_{1} & 0 & 0 & \cdots & 0 \\ 
0 & \phi_{2} &  0  & \cdots &0 \\ 
0 & 0 &  \phi_{3} &\cdots &0 \\ 
\vdots & \vdots &\vdots & \ddots   &\vdots & \\
0 & 0  & 0 & \cdots & \phi_{S} \\ 
\end{bmatrix}.
\end{equation}

{\bf Case 3 (VAR 2-NN)}

All these previous methods assume that an observation at location $3$ and time $3$ ($y_{3,3}$) is influenced by the previous observation in time $y_{3,2}$ at the same location, but not affected by other sites at previous time points (e.g. $y_{22}$).    
To overcome this limitation, we consider a third variation based on the K-Nearest Neighbours (K-NN), which has been used to account for spatial dependence. See for example the case of Gaussian processes based on nearest neighbours in \citet{datta2016hierarchical, finley2017applying}.
In our modelling framework, we establish temporal dependence in the stream network between a spatial location and its 2 Nearest Neigbours (2-NN) based on total hydrological or total stream distance. 

We formulate this variation of $\pmb{\Phi}$ for the example in Fig~\ref{fig:network} as follows:

\begin{equation}
\pmb{\Phi}_1 = \begin{bmatrix}
\phi_{11} &\phi_{12} & \phi_{13} & 0 \\ 
\phi_{21}  & \phi_{22} &  \phi_{23}   & 0 \\ 
\phi_{31}  & \phi_{32} &  \phi_{33} & 0  \\ 
0 & \phi_{42}   & \phi_{43}  & \phi_{44} \\ 
\end{bmatrix}.
\end{equation}

\noindent The columns in this matrix represent {\it from} and the rows {\it to} a spatial location.
Alternatively, if we consider only temporal dependence from locations that are upstream:    

\begin{equation}
\pmb{\Phi}_1 = \begin{bmatrix}
\phi_{11} &0 & 0 & 0 \\ 
0  & \phi_{22} &  0   & 0 \\ 
\phi_{31}  & \phi_{32} &  \phi_{33} & 0  \\ 
0 & \phi_{42}   & \phi_{43}  & \phi_{44} \\ 
\end{bmatrix}.
\end{equation}

\noindent In this last formulation, note that spatial locations 1 and 2 have no neighbours upstream.

\subsection{Bayesian hierarchical model}
\label{bhm}

We define the model hierarchically as follows:

\begin{equation}
[\pmb{y}_{t}\mid \pmb{y}_{t-1},\pmb{\theta},\pmb{X}_{t},\pmb{X}_{t-1},\pmb{\beta}, \pmb{\Phi}_1, \pmb{\Sigma}] = \mathcal{N}(\pmb{\mu}_{t}, \pmb{\Sigma} + \sigma^{2}_{0} \pmb{I}),
\end{equation}

\begin{equation}
\pmb{\mu}_{t} = \pmb{X}_{t}\pmb{\beta} + \pmb{\Phi}_1 (\pmb{y}_{t-1} - \pmb{X}_{t-1}\pmb{\beta}) 
\end{equation}

\noindent where $\pmb{\Sigma}$  can be expressed as a combination of the vectors of spatially structured random effects for tail-up, tail-down, and Euclidean models from Eq~\ref{eq:covs}, $\pmb{\Sigma} = COV(\pmb{u} + \pmb{d}+ \pmb{e})$, and

\begin{equation}
[\pmb{u} \mid \sigma^2_{u}, \alpha_{u}] = \mathcal{N}
(0, \sigma^2_{u} \pmb{R}(\alpha_{u})),
\label{eq:tum}
\end{equation}

\begin{equation}
[\pmb{d} \mid \sigma^2_{d}, \alpha_{d}] = \mathcal{N}
(0, \sigma^2_{d} \pmb{R}(\alpha_{d})),
\end{equation}

\begin{equation}
[\pmb{e} \mid \sigma^2_{e}, \alpha_{e}] = \mathcal{N}
(0, \sigma^2_{e} \pmb{R}(\alpha_{e})),
\label{eq:edm}
\end{equation}

with (hyper) priors (distributions that are given in \ref{sec:hierar})
\begin{equation}
[\pmb{\beta}][\sigma^2_0][\pmb{\Phi}_{1}][\sigma^2_{u}][\alpha_{u}][\sigma^2_{d}][\alpha_{d}][\sigma^2_{e}][\alpha_{e}].
\end{equation}

Then our hierarchical model yields the joint distribution

\begin{equation}
[\mathbf{y}, \mathbf{u}, \mathbf{d}, \mathbf{e}, \pmb{\beta}, \pmb{\Phi}_{1}, \sigma^2_{0}, \sigma^2_{u}, \alpha_{u}, \sigma^2_{d}, \alpha_{d}, \sigma^2_{e}, \alpha_{e}\mid \mathbf{X}],
\end{equation}

where the posterior distribution is proportional to this, so, using  MCMC, we obtain a sample from 

\begin{equation}
[\mathbf{u}, \mathbf{d}, \mathbf{e}, \pmb{\beta},  \pmb{\Phi}_{1}, \sigma^2_{0}, \sigma^2_{u}, \alpha_{u}, \sigma^2_{d}, \alpha_{d}, \sigma^2_{e}, \alpha_{e}\mid \mathbf{y}, \mathbf{X}].
\end{equation}

The full hierarchical model representation is given in \ref{sec:hierar}.

The matrix $\pmb{\Phi}$ can be formulated in Case 2 as follows:  

\begin{itemize}
    \item (2a) a uniform or a truncated normal prior on the site-specific autoregression parameter $\phi_s \sim \mathcal{U}(-1, 1)$ or $\phi_s \sim \mathcal{N}(0.5, 0.2)T[-1,1]$. Alternatively, a truncated normal prior on $\phi_s \sim \mathcal{N}(\mu_{\phi}, \sigma_{\phi})T[-1,1]$, where $\mu_{\phi}$ and $\sigma_{\phi}$ are the common hyperparameters with hyperpriors e.g.: $\mu_\phi \sim \mathcal{N}(0.5, 0.2)$ and $\sigma_\phi \sim U(0, 2)$.
    Because the $\phi$'s are site-specific, there is no obvious way to make predictions using simple kriging on new sites with this approach.
    
    \item (2b) expressing $\phi_s$ as a linear combination of some covariates,  \begin{equation}
        \textrm{logit}(\phi_s) = \gamma_0 + \gamma_1  X_{1s} + \gamma_2  X_{2s} + \cdots + \gamma_J  X_{JS}, 
        \label{eq:logit}
    \end{equation}
    \noindent where $\{\gamma_{j}\}$ are regression coefficients and $\{X_{js}\}$ are fixed covariate values.  
    This model allows the temporal dependence to be conditioned on characteristics that are location specific and do not vary through time such as elevation or watershed area.
    Alternatively, the logit transformation can be replaced by $(e^{x} - 1 ) / (e^{x} + 1)$ that will restrict $\phi_s$ to take values from -1 to 1.
    This model is prefered over (2a) as we discuss later when we want to make predictions in areas where no data are available.
\end{itemize}

\subsection{Model selection criteria}

Selecting the most suitable model is a complicated matter and the ultimate decision may be based on considering a range of statistical, computational and practical factors. 
Here we list some criteria which will be considered for model selection.

\begin{enumerate}
    \item Estimation/Prediction accuracy: Comparing models based on the out-of-sample prediction RMSPE or cross-validation is common when the main aim is to make predictions or impute missing data. 
    We create two partitions of the data into a training (80\%) and a testing set (20\%). We refer to the predictions made in the testing dataset as imputation, to avoid confusion with the predictions we make in the rest of the stream network in a second stage.
    We assume that the true temperature in the testing set is latent or non-observed. 
    The model will be fit using the training data and we will use the parameter estimates and covariates to estimate the latent temperature in the testing set for validation of the model. 
    Additionally, information criteria such as the 
    Widely Applicable Information Criterion (WAIC) and approximate leave-one-out cross-validation (LOO) are also frequently used to assess model prediction accuracy. 
We also measure the performance of the probabilistic estimates in the models using the testing set based on the Continuous Ranked Probability Score (CRPS) \citep{gneiting2005calibrated}. This score compares the cumulative distribution functions of the prediction points to the observations and the smaller the CRPS the better is the prediction.

From the application point of view, the aim could be to obtain precise estimates of the fixed effects. To do so, we can compare models according to the Monte Carlo Standard Error of the fixed effects mean rank (SE rank). 
We also use the coverage, which indicates the goodness of fit by comparing the 95\% posterior predicted distribution in the testing set to the true latent temperature. The coverage is calculated as the proportion of posterior intervals containing the true value. A suitable coverage would result in 95\% of the samples in the training set contained in the posterior density interval. 
Using the two-sided exact binomial test statistic, we assessed whether the proportion is significantly different from 0.95 at significance levels of 0.05 and 0.10 to create three categories for the coverage.

\item Computing time and memory requirements: Bayesian models that rely on Markov chain Monte Carlo (MCMC) to approximate the posterior distribution tend to be computationally intensive. This is accentuated even more for spatial and spatio-temporal models that involve inverting large covariance matrices.
A constrain for several applications is the computational time.
    
\item Parameter identifiability, parsimony and statistical interpretability: Bayesian models are often over-parametrized, i.e. having more parameters than what can be effectively estimated from the data. This often causes convergence issues and increased computing time. 
    In general, models with a fewer number of parameters and less complexity that produce suitable performance are usually preferred under the principle of parsimony. 
    Models whose outcomes and parameters can easily be understood, primarily by practitioners, are said to be more interpretable. 
    
\end{enumerate}

\section{Case study}

{ \scriptsize\emph{``Nothing remains the same from one moment to the next, you can't step into the same river twice. Life--evolution--the whole universe of space/time, matter/energy--existence itself--is essentially change.''} - Ursula K. Le Guin}

\vspace{.5cm}

Thermal regimes (i.e. spatio-temporal stream-temperature dynamics) are critically important in determining habitat suitability and population persistence in freshwater streams \citep{isaak2017norwest}, which are currently experiencing declines in biodiversity far greater than the most threatened terrestrial systems \citep{dudgeon2006freshwater}. As such, stream temperature is often used to measure habitat impairment \citep{todd2008development, US2003} and serves as the basis for regulatory actions \citep{olden2010incorporating, rivers2013towards}.

\begin{figure}[ht]
	\centering
		\includegraphics[width=6.5in]{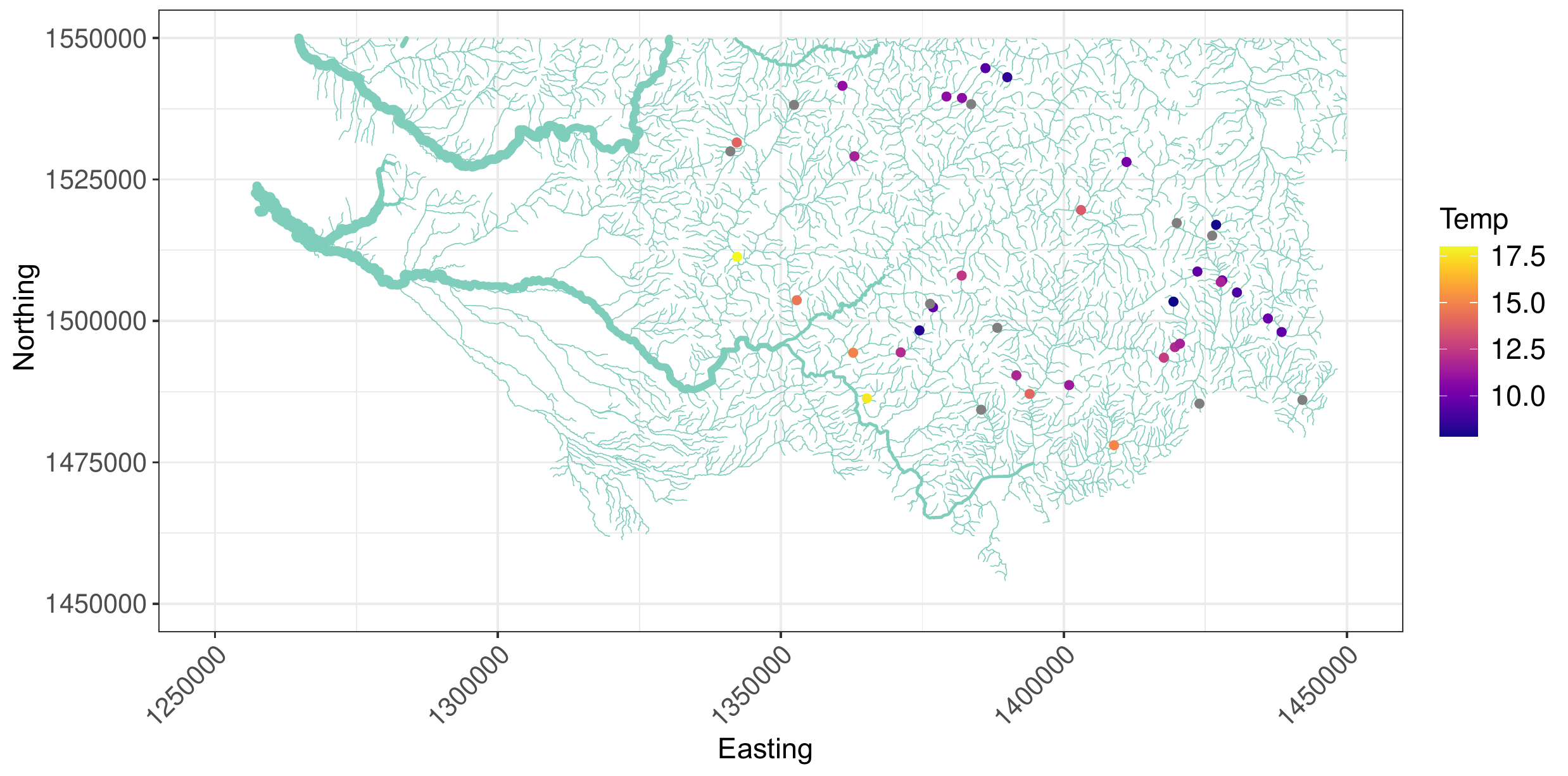}
	\caption{Extent of stream network used in the case study, along with daily mean temperature ($^\circ$C) values at 42 sensor sites on 2015-08-19. The width of the stream segments is proportional to the stream order. } 
	\label{fig:network2}
\end{figure}

In this case study, we use a water temperature dataset from the Boise River Basin, USA (Fig~\ref{fig:network2}), which is approximately 10,000 $\textrm{km}^2$ in size and includes 7,364 km of perennial streams. The basin provides designated critical habitat for bull trout (\textit{Salvelinus confluentus}), a threatened cold-water salmon species.  Temperature data were collected at 42 fixed spatial locations over a consecutive five-year period using in-situ sensors \citep{isaak2018principal} that recorded measurements at 30-minute intervals. These measurements were aggregated to generate daily mean temperatures, which we used as the response variable. In addition, 6422 prediction locations where the temperature was not observed were placed at 1km intervals on streams throughout the network to create a regular grid for mapping temperature model predictions. 
Stream temperature time series are zero bounded because water temperatures do not drop below freezing.
A spatial stream network (SSN) object was then created using the STARS software tool \citep{peterson2014stars}.

\subsubsection*{Time series}
\begin{figure}[ht]
	\centering
		\includegraphics[width=6.5in]{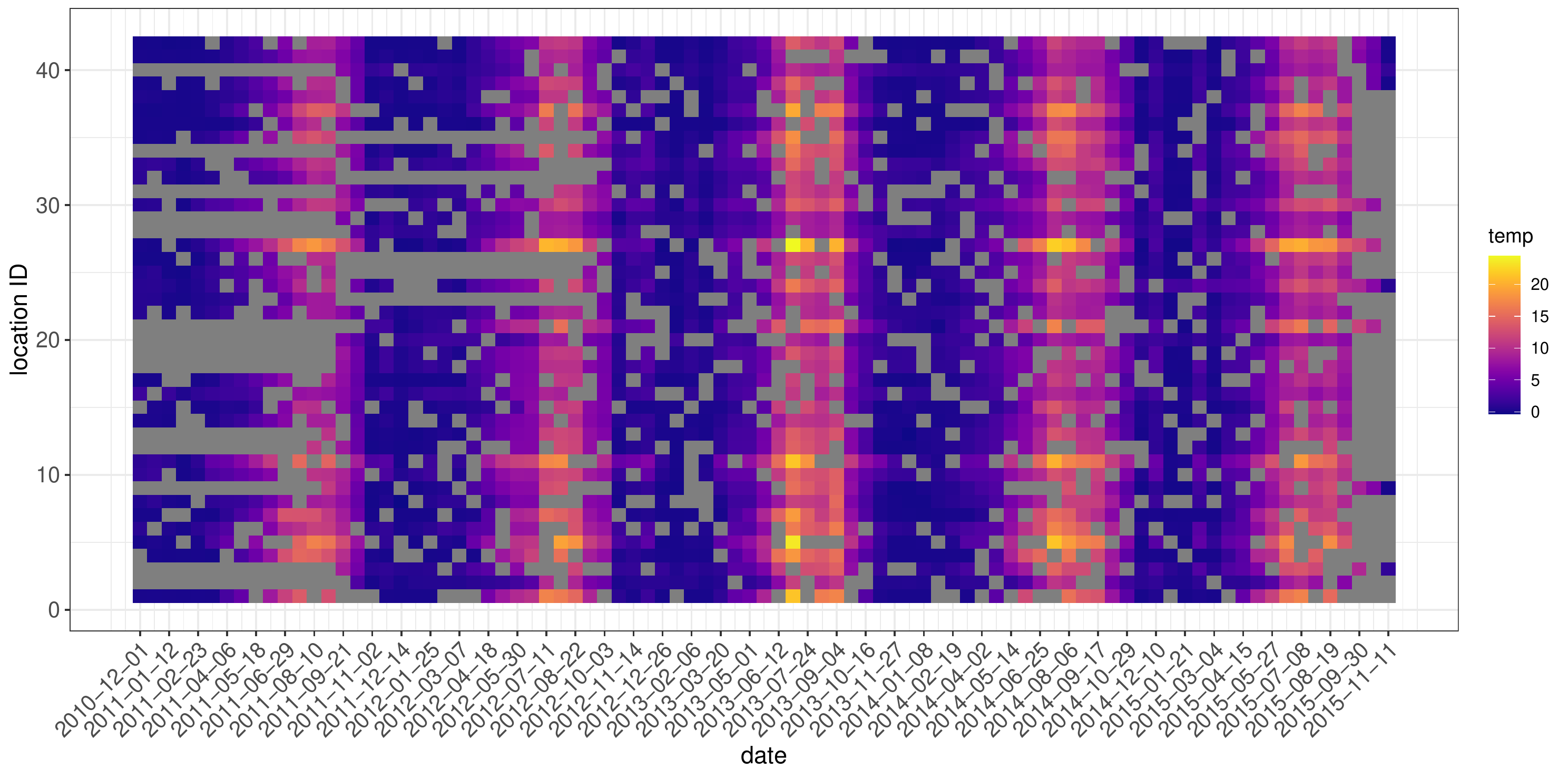}
	\caption{Time series of the daily mean temperature ($^\circ$C) values from the 42 spatial locations. The gray areas represent values of water temperature that are part of the testing set or they are missing values. } 
	\label{fig:time_series}
\end{figure}

Water temperature was measured from 2010-12-01 to 2015-12-01 at each of the 42 spatial locations. However, we subsampled the time series systematically with a frequency equal to 21 days for illustration purposes of the case study. This produced a longitudinal dataset consisting of 87 dates, which we used for all further analyses (Fig.\ref{fig:time_series}).
Note that some temperature values are missing in the original dataset due to sensor issues.
To assess the out-of-sample prediction accuracy of the models, we also randomly split the dataset, with 80\% set aside for model training and the remaining 20\% (gray areas in Fig.\ref{fig:time_series}) used for testing.
One of the benefits of implementing these models using a Bayesian framework is that missing values in the response variable (water temperatures values in the training set) are imputed on the go. 

\subsubsection*{Covariates}

Stream temperature is affected by a variety of topographic and climatic factors, such as stream slope, elevation and watershed area \citep{isaak2017norwest}, as well as air temperature \citep[e.g.][]{bal2014hierarchical}. Topographic covariates were obtained from the National Stream Internet dataset \citep{nagel2015national} and mean air temperature estimates were obtained from the  Parameter-elevation Regressions on Independent Slopes Model (PRISM) website (\url{https://prism.oregonstate.edu}).
The air temperature estimates were provided as raster datasets, with a 4km spatial resolution. Daily values for the 87 stream temperature dates were extracted at the observation and prediction locations on the Boise River network using the \texttt{raster} package in R \citep{raster}.

\begin{figure}[ht]
	\centering
	\includegraphics[width=7.5in]{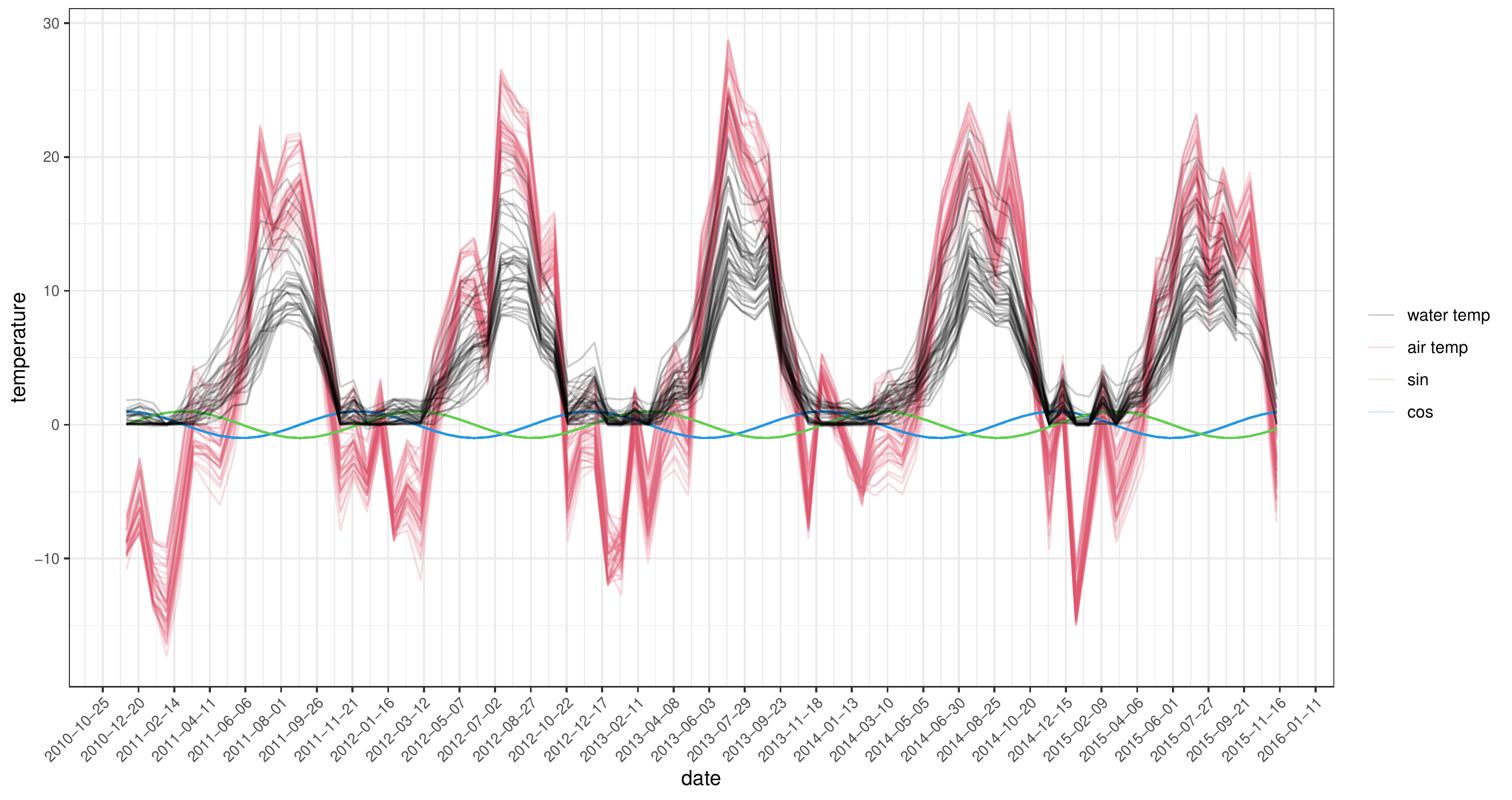}
	\caption{Dynamic of the water and air temperatures (in $^\circ$C) across the 87 dates (covering 5 years). 42 time series for each (water and air temperature) are shown, representing the spatial locations. 
	The first two Fourier terms are also included in green and blue. } 
    \label{fig:water_air_sin_cos}
\end{figure}

Annual seasonality in water temperature across years are often modelled using harmonic covariates or Fourier terms \citep[e.g.][]{wikle1998hierarchical, bal2014hierarchical, graf2018distribution}.
In our case, this helps account for seasonal and interannual changes in the air temperature-water temperature relationship that occurs in snowmelt-dependent systems like our case history area. 
The first pair of Fourier terms ($\textrm{sin}$ and $\textrm{cos}$) were obtained using the function Fourier from the R package \texttt{forecast} \citep{forecast}:

\begin{equation}
\textrm{sin}_t = \sin(\frac{2\pi t}{m}),
\end{equation}

\begin{equation}
\textrm{cos}_t = \cos(\frac{2\pi t}{m}),
\end{equation}

\noindent where $m = 365$ and $t$ is the time period. Note that this produces identical values for all spatial locations at time $t$ since these are temporal covariates.
Fig.\ref{fig:water_air_sin_cos} shows the water and air temperatures time series along with the first pair of Fourier terms ($\sin$ and $\cos$).

The fixed effects part in Eq~\ref{eq:err_} is then:

\begin{equation}
\pmb{X}_{t}\pmb{\beta} = \beta_0 + \beta_1 \times \textrm{slope} + \beta_2 \times \textrm{elev}  + \beta_3 \times \textrm{cumdrain} + \beta_4 \times \textrm{airtemp}_{t} +  \beta_5 \times \textrm{sin}_{t} + \beta_6 \times \textrm{cos}_{t},
\end{equation}

\noindent where $\textrm{slope}$, $\textrm{elev}$ and $\textrm{cumdrain}$  are vectors of dimension $S \times T$ corresponding to the slope, elevation and the cumulative drainage area at the sampling sites and they are constant across all the time points.
The $\textrm{airtemp}$ is a $S \times T$ vector of air temperature values at each of the spatial locations at each of the time points.

For the Case 2b formulation, we included covariates that are expected to affect the temporal autocorrelation at a location, in this case the standardized elevation (elev) and watershed area (ws) variables were used to estimate $\pmb{\Phi}$, where:

    \begin{equation}
   \textrm{logit}(\phi_s) = \beta_0 + \beta_1 \times \textrm{elev} + \beta_2 \times \textrm{ws}.
        \label{eq:logit2}
    \end{equation}
    
Note that we did not detect substantial multicollinearity between elevation (elev) and watershed area (ws). 
Interestingly, the watershed area could be used in the models in three different ways: (I) as a fixed effect in the linear part of the regression formulation, (II) as the basis of the AFV used in the tail-up covariance structure and (III) as a covariate for determining $\phi_s$ in Case 2b.

We fit several model variations using different spatial covariance matrices ($\pmb{\Sigma}$ from Eq~\ref{eq:covs}) and AR/VAR variations. 
Each model contained one AR/VAR structure (AR, VAR(1), VAR\_2b or VAR\_2NN) and one or two spatial components \{tail-down (td), tail-up (tu), Euclidean distance (ed) or the combination of two of them: tutd, tued or tded\}.
This resulted in four groups of models with different temporal structures, with each group composed of six spatial variations. See Table~\ref{table:methodstab}.
The Euclidean distance between spatial locations was obtained from the x-coordinate and y-coordinate in the data represented using an Albers projection.
In the implementation, the MCMC draws from the components  $u$, $t$ and $e$ of the spatial process in Eq~\ref{eq:covs} were common for all the time points. 
Computations were performed on a High-Performance Computing (HPC) system. 

We compared the models using six prediction/estimation outcomes: the Widely Applicable Information Criterion (WAIC), leave-one-out cross-validation (LOO),
Continuous Ranked Probability Score (CRPS),
out-of-sample Root Mean Square Prediction Error (RMSPE) obtained using the testing dataset, Monte Carlo standard error of the fixed effects mean rank (SE rank) and the 95\% prediction coverage based on the testing set. The SE rank was computed ranking the posterior standard error of the seven fixed effects included in the model.

We also used three qualitative (subjective) factors to compare models: parameter identifiability, model complexity and model interpretability.
Finally, we compared models based on computing time and memory usage.

\subsection{Results of the case study}
\label{sec:Res}

\begin{figure}[ht]
	\centering
	\includegraphics[width=7.25in]{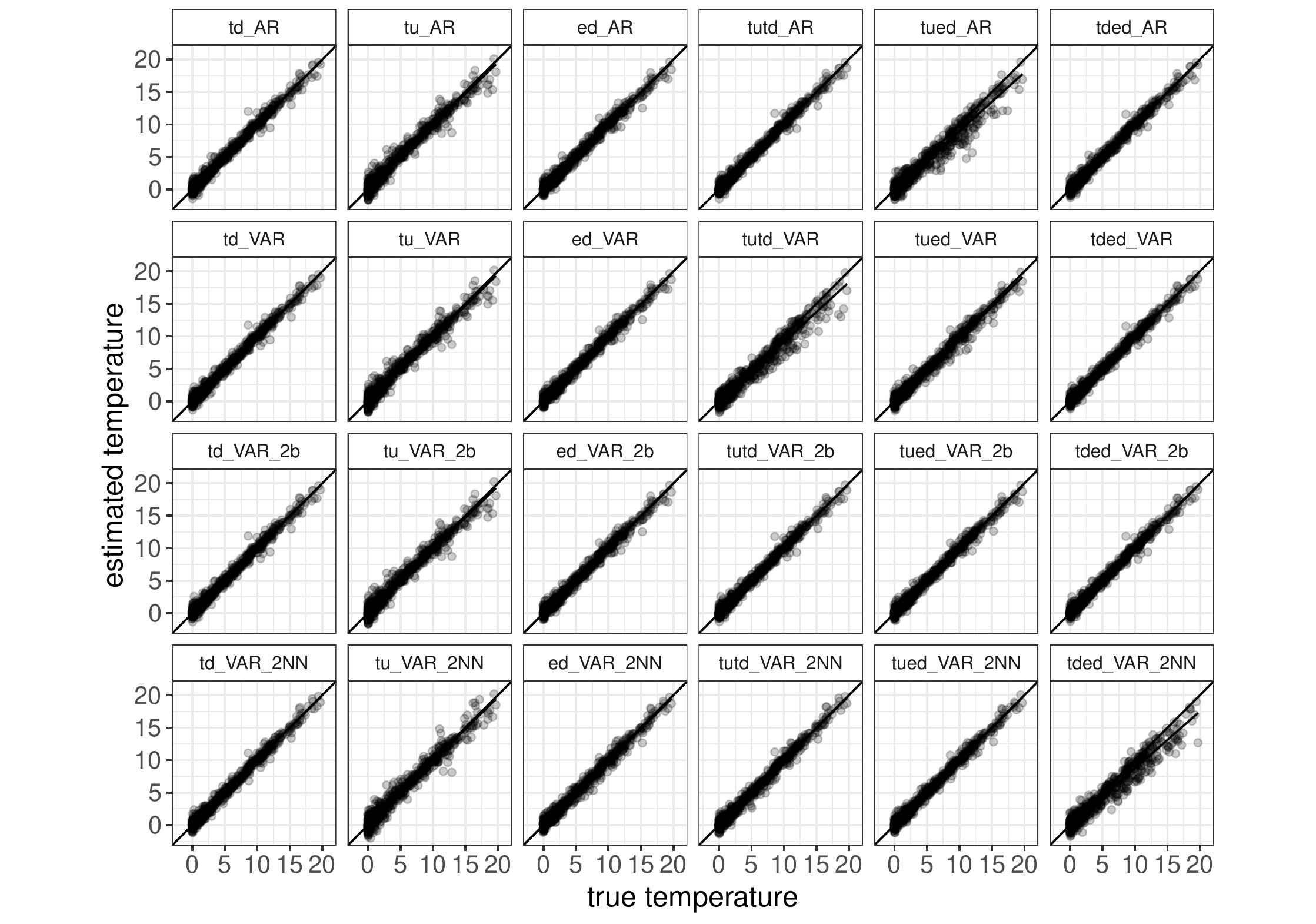}
	\caption{Comparison of the predicted (posterior mean) versus the true latent stream temperature values from the 24 models for the 87 dates used in the analysis.} 
    \label{fig:temp_pred_vs_obs}
\end{figure}

The results of the model comparison are shown in Table \ref{table:compar}, except memory usage, which did not vary considerably between models ($\approx$ 3.5Gb). The results showed that VAR\_2b and AR model variations had the smallest  WAIC and LOO values and that with a few exceptions, the RMSPE values were small (0.5-0.6 $^\circ$C) relative to the magnitude of the temperature values. Fig.\ref{fig:temp_pred_vs_obs} 
shows that the majority of models retrieved the true latent temperature in the testing dataset (hold out data) well and produced very accurate predictions. Although, from Table~\ref{table:compar} models that included the tail-down component appear to have a better predictive performance within each group.
Fig.\ref{fig:temp_pred_vs_obs_HDI} in the Appendix shows the 95\% highest density intervals (HDI) of the predictions in each of the models.

\begin{table}[]
\caption{Model combinations used in the case study }
\label{table:methodstab} 
\scalebox{0.76}{
\begin{tabular}{lllllll}
                   & \multicolumn{6}{l}{Spatial structure}  \\\hline 
Temporal structure & Tail-Down    & Tail-up      & Euc dist     & Tail-up/Tail-Down & Tail-up/Euc dist & Tail-down/Euc dist \\
AR (Case 1)        & td\_AR       & tu\_AR       & ed\_AR       & tutd\_AR          & tued\_AR         & tded\_AR           \\
VAR(1) (Case 2a)   & td\_VAR      & tu\_VAR      & ed\_VAR      & tutd\_VAR         & tued\_VAR        & tded\_VAR          \\
VAR(1) (Case 2b)   & td\_VAR\_2d  & tu\_VAR\_2d  & ed\_VAR\_2d  & tutd\_VAR\_2d     & tued\_VAR\_2d    & tded\_VAR\_2d      \\
VAR(1) NN (Case 3) & td\_VAR\_2NN & tu\_VAR\_2NN & ed\_VAR\_2NN & tutd\_VAR\_2NN    & tued\_VAR\_2NN   & tded\_VAR\_2NN   \\
\hline 
\end{tabular}
}
\end{table}

Although the predictive ability of most models was similar, there were some differences. For example, the SE rank which measures the uncertainty around the fixed effects favored the AR models, while the best prediction coverage was produced by the VAR models. 
The VAR\_2b model variations showed the best CRPS values.
Identifiability tended to be an issue in models that included multiple covariance structures (e.g. tutd, tued and tded), since they have two partial sill and spatial range parameters to estimate. We also assigned a higher complexity to the VAR\_2b and VAR\_NN models as we explained in the methods section. 
VAR\_2b models had better interpretability because they explain the amount of autoregression based on topological features. In contrast, interpretability was deemed more limited in the VAR\_NN, as well as AR and VAR models with combinations of spatial structures due to higher complexity.  

Nevertheless, some of the results (from Table \ref{table:compar}) should be interpreted with caution. First, not all the qualitative factors have the same importance and they might not be essential in many cases. In addition, a thorough assessment of computing time would require a more extensive experiment.

\begin{table}[ht]
\caption{Comparison of performance statistics for models fit to stream temperature dataset. The * symbol indicates that the column has been color-coded based on terciles. The categories in the qualitative factors are subjective, obtained based on our perception of identifiability, model complexity, and interpretability.} 
\label{table:compar}
\scalebox{0.85}{
\begin{tabular}{p{2.65cm}R{1.4cm}R{1.4cm}R{1.4cm}R{1.4cm}R{1.4cm}R{1.4cm}C{1.4cm}C{1.4cm}C{1.4cm}C{1.4cm}} 
 & \multicolumn{6}{c}{Prediction accuracy/Estimation}    & \multicolumn{1}{c}{Comp}  & \multicolumn{3}{c}{Qualitative factors} \\ 
 model & WAIC* & LOO* & CRPS* & RMSPE* & SE rank* & cover &time* & ident & complex & interp \\ 
  \hline
 td\_AR & \cellcolor{ao}9098 & \cellcolor{ao}9170 & \cellcolor{carrotorange}0.289& \cellcolor{ao}0.549 & \cellcolor{ao}2 & \cellcolor{ao}0.942 & \cellcolor{bred}hig & \cellcolor{ao}good & \cellcolor{ao}low & \cellcolor{ao}good \\ 
tu\_AR & \cellcolor{carrotorange}11874 & \cellcolor{carrotorange}11934 &\cellcolor{bred}0.439& \cellcolor{bred}0.844 & \cellcolor{ao}1 & \cellcolor{bred}0.969 & \cellcolor{bred}hig & \cellcolor{ao}good & \cellcolor{ao}low & \cellcolor{ao}good \\ 
 ed\_AR & \cellcolor{carrotorange}10204 & \cellcolor{carrotorange}10355 &\cellcolor{carrotorange}0.314& \cellcolor{carrotorange}0.576 & \cellcolor{ao}5 & \cellcolor{ao}0.963 & \cellcolor{ao}low & \cellcolor{ao}good & \cellcolor{ao}low & \cellcolor{ao}good \\ 
  tutd\_AR & \cellcolor{ao}8840 & \cellcolor{ao}8901 &\cellcolor{ao}0.277& \cellcolor{ao}0.53 & \cellcolor{ao}4 & \cellcolor{ao}0.947 & \cellcolor{bred}hig & \cellcolor{carrotorange}mod & \cellcolor{ao}low & \cellcolor{carrotorange}mod \\ 
  tued\_AR & \cellcolor{bred}64540 & \cellcolor{bred}15371 &\cellcolor{bred}0.602& \cellcolor{bred}1.241 & \cellcolor{carrotorange}13 & \cellcolor{bred}0.985 & \cellcolor{ao}low & \cellcolor{carrotorange}mod & \cellcolor{ao}low & \cellcolor{carrotorange}mod \\ 
  tded\_AR & \cellcolor{ao}9024 & \cellcolor{ao}9210 &\cellcolor{carrotorange}0.299& \cellcolor{carrotorange}0.565 & \cellcolor{carrotorange}8 & \cellcolor{ao}0.947 & \cellcolor{carrotorange}med & \cellcolor{carrotorange}mod & \cellcolor{ao}low & \cellcolor{carrotorange}mod \\ \hline  
  td\_VAR & \cellcolor{carrotorange}10013 & \cellcolor{carrotorange}9518 &\cellcolor{carrotorange}0.286 &\cellcolor{ao}0.54 & \cellcolor{bred}15 & \cellcolor{ao}0.957 & \cellcolor{bred}hig & \cellcolor{ao}good & \cellcolor{ao}low & \cellcolor{ao}good \\ 
  tu\_VAR & \cellcolor{bred}11890 & \cellcolor{bred}11959 &\cellcolor{bred}0.447& \cellcolor{bred}0.854 & \cellcolor{ao}2 & \cellcolor{carrotorange}0.965 & \cellcolor{bred}hig & \cellcolor{ao}good & \cellcolor{ao}low & \cellcolor{ao}good \\ 
  ed\_VAR & \cellcolor{bred}27476 & \cellcolor{bred}14838 &\cellcolor{carrotorange}0.304& \cellcolor{bred}2.446 & \cellcolor{bred}19 & \cellcolor{bred}0.977 & \cellcolor{ao}low & \cellcolor{ao}good & \cellcolor{ao}low & \cellcolor{ao}good \\ 
  tutd\_VAR & \cellcolor{bred}73033 & \cellcolor{bred}14225 & \cellcolor{bred}0.52  & \cellcolor{bred}1.109 & \cellcolor{bred}16 & \cellcolor{bred}0.971 & \cellcolor{ao}low & \cellcolor{carrotorange}mod & \cellcolor{ao}low & \cellcolor{carrotorange}mod \\ 
  tued\_VAR & \cellcolor{bred}19829 & \cellcolor{bred}13196 & \cellcolor{bred}0.352 & \cellcolor{carrotorange}0.676 &\cellcolor{carrotorange}10 & \cellcolor{bred}0.968 & \cellcolor{carrotorange}med & \cellcolor{carrotorange}mod & \cellcolor{ao}low & \cellcolor{carrotorange}mod \\ 
  tded\_VAR & \cellcolor{carrotorange}10373 & \cellcolor{carrotorange}10145 &  \cellcolor{ao}0.28 & \cellcolor{ao}0.527 &  \cellcolor{bred}14 & \cellcolor{carrotorange}0.965 & \cellcolor{carrotorange}med & \cellcolor{carrotorange}mod & \cellcolor{ao}low & \cellcolor{carrotorange}mod \\ \hline  
  td\_VAR\_2b & \cellcolor{ao}8832 & \cellcolor{ao}8950 &\cellcolor{ao}0.285& \cellcolor{carrotorange}0.55 & \cellcolor{carrotorange}6 & \cellcolor{ao}0.945 & \cellcolor{carrotorange}med & \cellcolor{ao}good & \cellcolor{carrotorange}mod & \cellcolor{ao}good \\ 
  tu\_VAR\_2b & \cellcolor{carrotorange}11870 & \cellcolor{bred}11997 &\cellcolor{bred}0.439& \cellcolor{carrotorange}0.843 & \cellcolor{ao}4 & \cellcolor{bred}0.968 & \cellcolor{bred}hig & \cellcolor{ao}good & \cellcolor{carrotorange}mod & \cellcolor{ao}good \\ 
  ed\_VAR\_2b & \cellcolor{carrotorange}10000 & \cellcolor{carrotorange}9977 &\cellcolor{carrotorange}0.309& \cellcolor{carrotorange}0.567 & \cellcolor{bred}17 & \cellcolor{ao}0.957 & \cellcolor{carrotorange}med & \cellcolor{ao}good & \cellcolor{carrotorange}mod & \cellcolor{ao}good \\ 
  tutd\_VAR\_2b & \cellcolor{ao}8290 & \cellcolor{ao}8399 &\cellcolor{ao}0.278& \cellcolor{ao}0.537 & \cellcolor{carrotorange}7 & \cellcolor{ao}0.939 & \cellcolor{carrotorange}med & \cellcolor{carrotorange}mod & \cellcolor{carrotorange}mod & \cellcolor{ao}good \\ 
  tued\_VAR\_2b & \cellcolor{ao}9248 & \cellcolor{ao}9132 &\cellcolor{ao}0.282& \cellcolor{ao}0.534 & \cellcolor{bred}17 & \cellcolor{ao}0.96 & \cellcolor{ao}low & \cellcolor{carrotorange}mod & \cellcolor{carrotorange}mod & \cellcolor{ao}good \\ 
  tded\_VAR\_2b & \cellcolor{ao}9248 & \cellcolor{ao}9132 &\cellcolor{ao}0.281& \cellcolor{ao}0.534 & \cellcolor{bred}18 & \cellcolor{ao}0.942 & \cellcolor{ao}low & \cellcolor{carrotorange}mod & \cellcolor{carrotorange}mod & \cellcolor{ao}good \\ \hline  
  td\_VAR\_2NN & \cellcolor{ao}8755 & \cellcolor{ao}8914 &\cellcolor{ao}0.282& \cellcolor{ao}0.518 & \cellcolor{ao}5 & \cellcolor{bred}0.933 & \cellcolor{bred}hig & \cellcolor{carrotorange}mod & \cellcolor{carrotorange}mod & \cellcolor{carrotorange}mod \\ 
  tu\_VAR\_2NN & \cellcolor{carrotorange}11426 & \cellcolor{carrotorange}11478 &\cellcolor{bred}0.462& \cellcolor{bred}0.886 & \cellcolor{ao}3 & \cellcolor{ao}0.948 & \cellcolor{bred}hig & \cellcolor{carrotorange}mod & \cellcolor{carrotorange}mod & \cellcolor{carrotorange}mod \\ 
  ed\_VAR\_2NN & \cellcolor{carrotorange}11295 & \cellcolor{carrotorange}10364 &\cellcolor{carrotorange}0.325& \cellcolor{carrotorange}0.595 & \cellcolor{carrotorange}9 & \cellcolor{ao}0.957 & \cellcolor{ao}low & \cellcolor{carrotorange}mod & \cellcolor{carrotorange}mod & \cellcolor{carrotorange}mod \\ 
  tutd\_VAR\_2NN & \cellcolor{bred}13212 & \cellcolor{carrotorange}9485 &\cellcolor{carrotorange}0.305& \cellcolor{carrotorange}0.584 & \cellcolor{carrotorange}12 & \cellcolor{bred}0.928 & \cellcolor{carrotorange}med & \cellcolor{carrotorange}mod & \cellcolor{carrotorange}mod & \cellcolor{carrotorange}mod \\ 
  tued\_VAR\_2NN & \cellcolor{bred}$>$90000 & \cellcolor{bred}78866 &\cellcolor{ao}0.28& \cellcolor{bred}2.593 & \cellcolor{bred}20 & \cellcolor{ao}0.937 & \cellcolor{ao}low & \cellcolor{carrotorange}mod & \cellcolor{carrotorange}mod & \cellcolor{carrotorange}mod \\ 
  tded\_VAR\_2NN & \cellcolor{bred}82104 & \cellcolor{bred}14774 &\cellcolor{bred}0.615& \cellcolor{bred}1.327 & \cellcolor{carrotorange}11 & \cellcolor{bred}0.969 & \cellcolor{carrotorange}med & \cellcolor{carrotorange}mod & \cellcolor{carrotorange}mod & \cellcolor{carrotorange}mod \\ 
   \hline         
\end{tabular} 
}
\end{table}

\clearpage

\subsection*{Results from the td\_VAR\_2b model }

\begin{figure}[ht]
\begin{subfigure}{1\textwidth}
\centering
\caption{} 
   \includegraphics[width=7.5in]{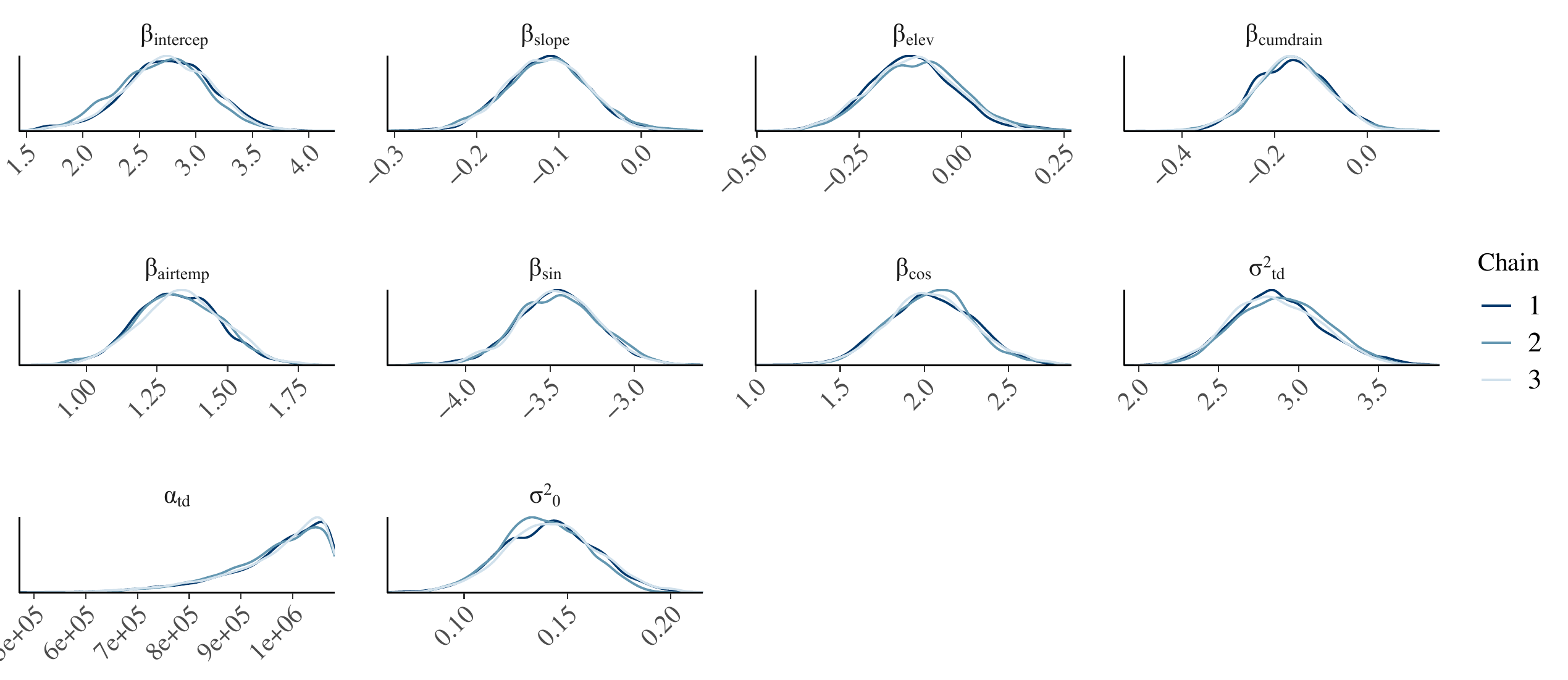} 
   \label{fig:VAR_post_A} 
\end{subfigure}
\centering
\begin{subfigure}{1\textwidth}
 \centering
	\caption{}
  \includegraphics[width=6.5in]{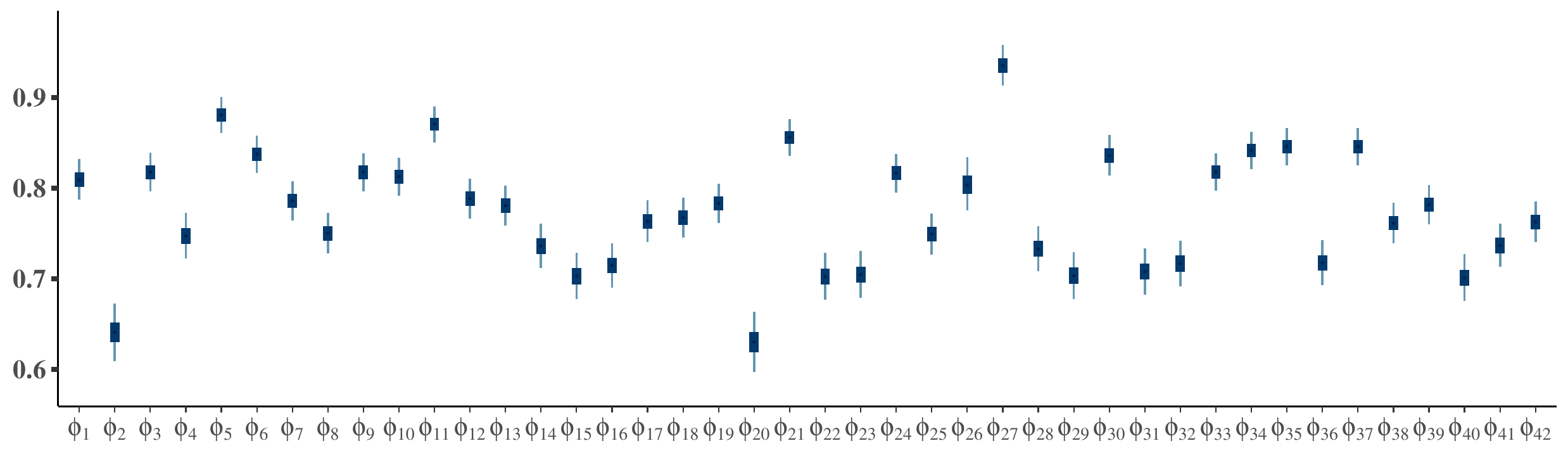}
  \label{fig:VAR_post_B}
\end{subfigure}%
\caption{Posterior distributions of $\beta$, $\sigma^2_{td}$ and $\alpha_{td}$ (a) and $\pmb{\Phi}$ (b).}
\end{figure}

In this section, we discuss the main results from the td\_VAR\_2b model from Table~\ref{table:compar}, that produced the second best WAIC/LOO after the tu\_td\_VAR\_2b model, but only involves one spatial structure.
Fig~\ref{fig:VAR_post_A} shows the posterior distributions of the parameters of interest. 
First, we note that the fixed effects (slope, cumulative drainage area and air temperature) and harmonic covariates have posterior distributions significantly different from zero, which suggests that they significantly affect stream temperature. 
The median of the spatial range $\alpha_{td}$ was approximately $10^6$ meters (1,000 km), which is a considerable range. This indicates that spatial autocorrelation exists between locations that are less than 1,000 km apart.
As expected, there is a high temporal dependence, with $\phi$ values ranging from 0.6 to 0.9 (Fig~\ref{fig:VAR_post_B}). 
We measured the ratio of percentage of spatially structured to independent residual variation that is explained by the model after accounting for the fixed effects - $\sigma_{td} / (\sigma_{td} + \sigma_{0}$) = 95.33\%.
This high proportion of spatial variation suggests that other covariates could be included to increase the prediction accuracy of the model.  
Fig.\ref{fig:perc_var_explained} shows a histogram of the posterior distribution of this ratio.

\begin{figure}[ht]
	\centering
	\includegraphics[width=3.5in]{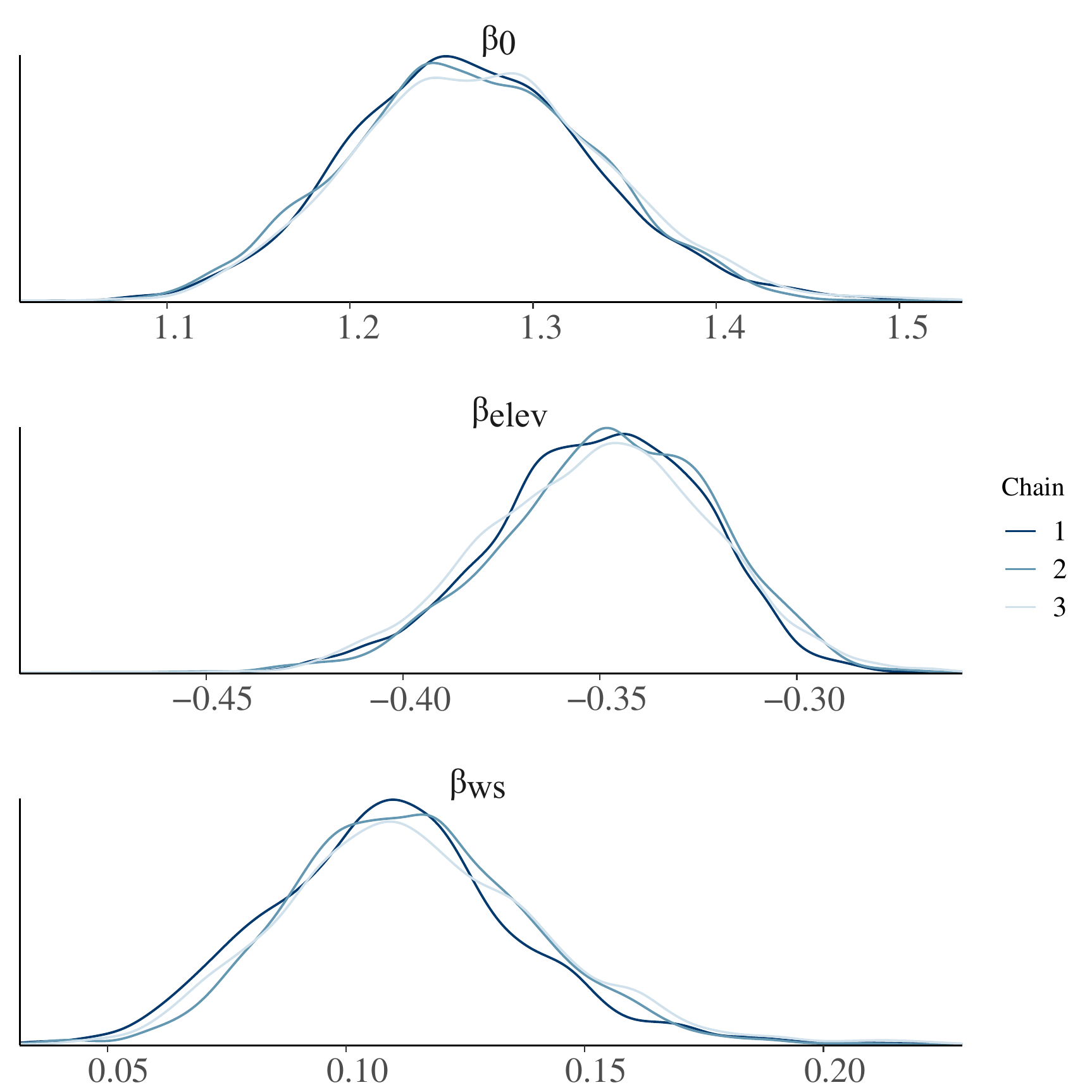}
	\caption{Posterior distribution of the regression coefficients associated with the elevation and watershed area affecting $\pmb{\Phi}$ in Eq~\ref{eq:logit} Case 2b.} 
    \label{fig:reg_coef_phi_distrib_VAR1_c2}
\end{figure}

Fig~\ref{fig:reg_coef_phi_distrib_VAR1_c2} shows the posterior regression coefficients associated with the standardised elevation and watershed area.
Both covariates substantially affected the amount of temporal dependence, with the amount of autoregression or temporal dependence decreasing with elevation. In addition, the segments in the downstream portion of the network with the larger watershed area also had larger values of $\phi$. This makes sense from a physical perspective because there is less thermal variability in large streams due to the inertial mass of water compared to small streams higher in the network.   

\subsection{Making predictions at unobserved locations}

In the previous section, we made predictions (imputation of missing values) on a testing dataset.
In this section, we are interested in making predictions across the whole network where observations have not occurred, borrowing strength from the observation sites in space and time. 

We produced a prediction dataset at the 6422 prediction locations regularly spaced every 1 km.
As shown in \ref{sec:sim} there are two main options for making predictions: (I) constructing the full space-time covariance matrix or (II) using the vector autoregression of the spatial model. 

Here we use the first option, known as the simple kriging approach is formulated as follows:

\begin{equation}
\widehat{\pmb{y}}_{P} = \pmb{X}_{P}\pmb{\beta} + {c_{OP}}'C^{-1}_{OO} (\pmb{y}_{O} - \pmb{X}_{O}\pmb{\beta} ), \\
    \label{eq:krig}
\end{equation}

\noindent where $O$ and $P$ refer to observation and prediction locations, respectively.
Here, $\widehat{\pmb{y}}_{P}$ is a stacked vector of predictions at the locations $P$ across all the time points $T$, while $\pmb{y}_{O}$  is a stacked vector with all the observations across the time points. 

The matrices $\pmb{X}_{P}$ and $\pmb{X}_{O}$ are space-time design matrices of covariates, and $\pmb{\beta}$ is a vector of (known) regression coefficients.  
Here, $C_{OO}$ is the covariance matrix of 
dimension $O \times T$ by $O \times T$,
between observation points defined in Eq~\ref{eq:covs} at different time points.
The covariance matrix $c_{OP}$ of dimension $O \times T$ by $P \times T$ between observation and prediction points will have the same structure as $C_{OO}$. That is, if $C_{OO}$ was obtained from an AR exponential tail-down model with parameters $\phi$, $\sigma_{td}$ and $\alpha_{td}$, we will use these parameters to construct $c_{OP}$.
Any missing values for observed sites across the time points will be imputed first using MCMC, which will give us a complete $O \times T$ data set.

Eq.~\ref{eq:krig} involves inverting the covariance matrix of the observations ($C_{OO}$), which can be obtained using: 
$$C^{-1}_{OO} = \pmb{\Sigma}_{OO}^{-1} \otimes \pmb{\Sigma}_{var}^{-1},$$
\noindent where $\pmb{\Sigma}_{OO}$ is the spatial covariance matrix defined in Eq~\ref{eq:covs} and $\pmb{\Sigma}_{var}$ is the temporal covariance matrix of the VAR(1) process.

See more details of these covariance matrices in \ref{sec:sim} and e.g. in  \citet{wikle2019spatio}.

In this application, the predictions were made using a tail-down autoregressive process (td-AR).
We use the MCMC stacked chains from the parameters in the fitted model object ($\pmb{\beta}$, $\sigma_{TD}$, $\alpha$,  $\sigma_0$ and $\phi$) to produce predictions in batches in parallel, based on the covariate values for these locations (slope, elevation, watershed area, air temperature and the first harmonic pair). 
The computations took approximately 3 hours.

\begin{figure}[ht]
	\centering
	\includegraphics[width=7.5in]{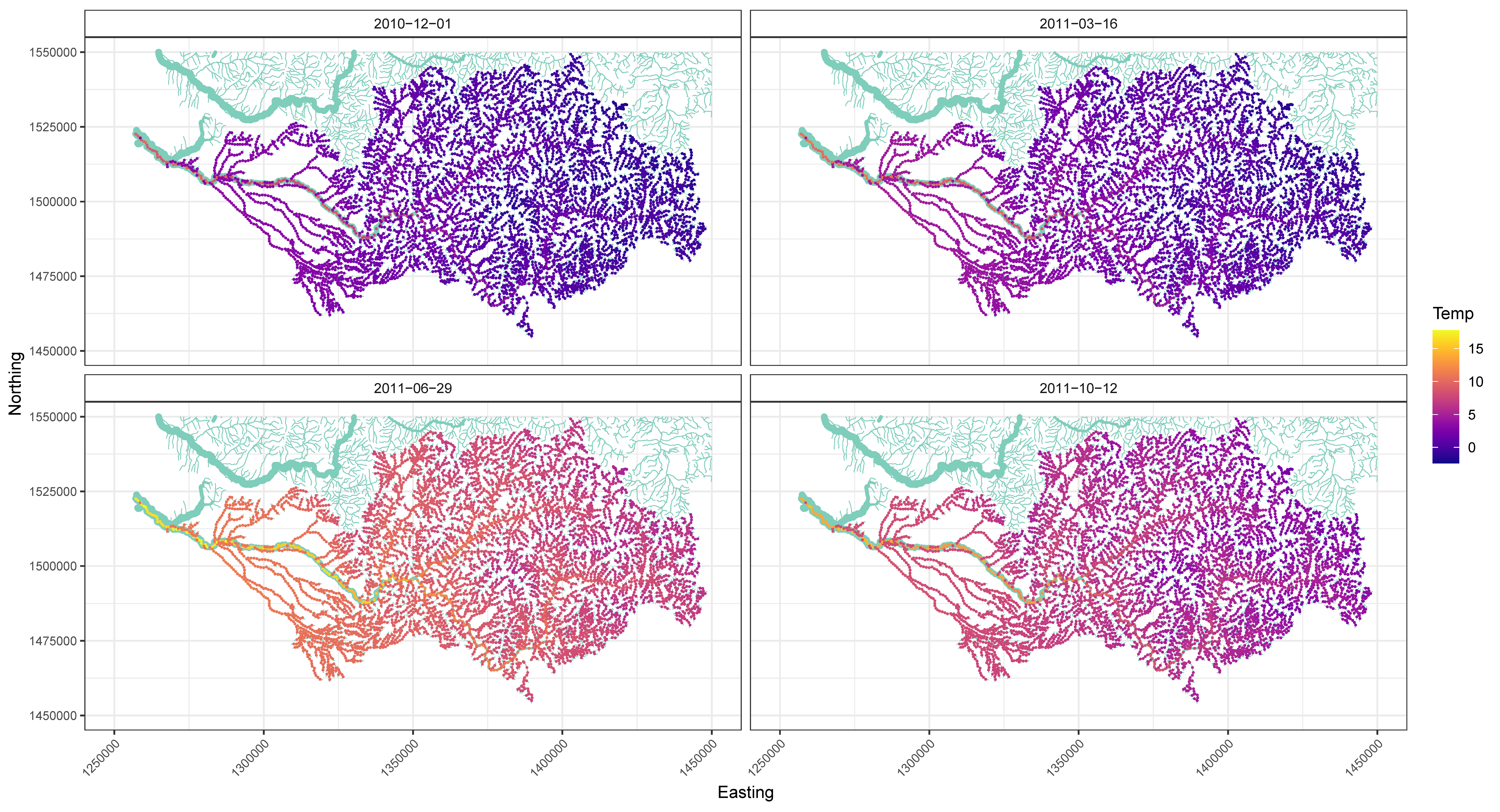}
	\caption{Posterior of the mean daily temperature in the Boise River Basin on four dates. } 
    \label{fig:mean_temp_boise_reduced}
\end{figure}

Fig~\ref{fig:mean_temp_boise_reduced} shows the posterior mean temperature throughout the Boise River Basin on four dates.  
Similarly, the model produces uncertainty around these estimates (not shown here). 

\subsubsection*{Exceedance probability}

In stream networks, regulatory statutes are often set based on the likelihood that the variable of interest will exceed a specified threshold \citep{money2009modern}. Where exceedence is common, management and restoration efforts may be targeted in efforts to decrease negative ecological impacts. In the case of water temperature, such efforts may involve timing of cold-water releases from upstream dams, limits on recreational and commercial fishing during thermally stressful periods for fish, or planting trees along stream banks to provide shade and lower temperatures. 
In this section, we identify stream segments with high chances of exceeding a critical thermal threshold for bull trout. This species has an especially cold thermal niche and the juveniles rarely inhabit where daily summer temperatures are higher than 13 $^\circ$C \citep{isaak2017big}. Temperatures in excess of this limit may either cause mortality or behavioral thermoregulation involving the movement of individuals to cooler areas.

\begin{figure}[ht]
	\centering
	\includegraphics[width=8.75in]{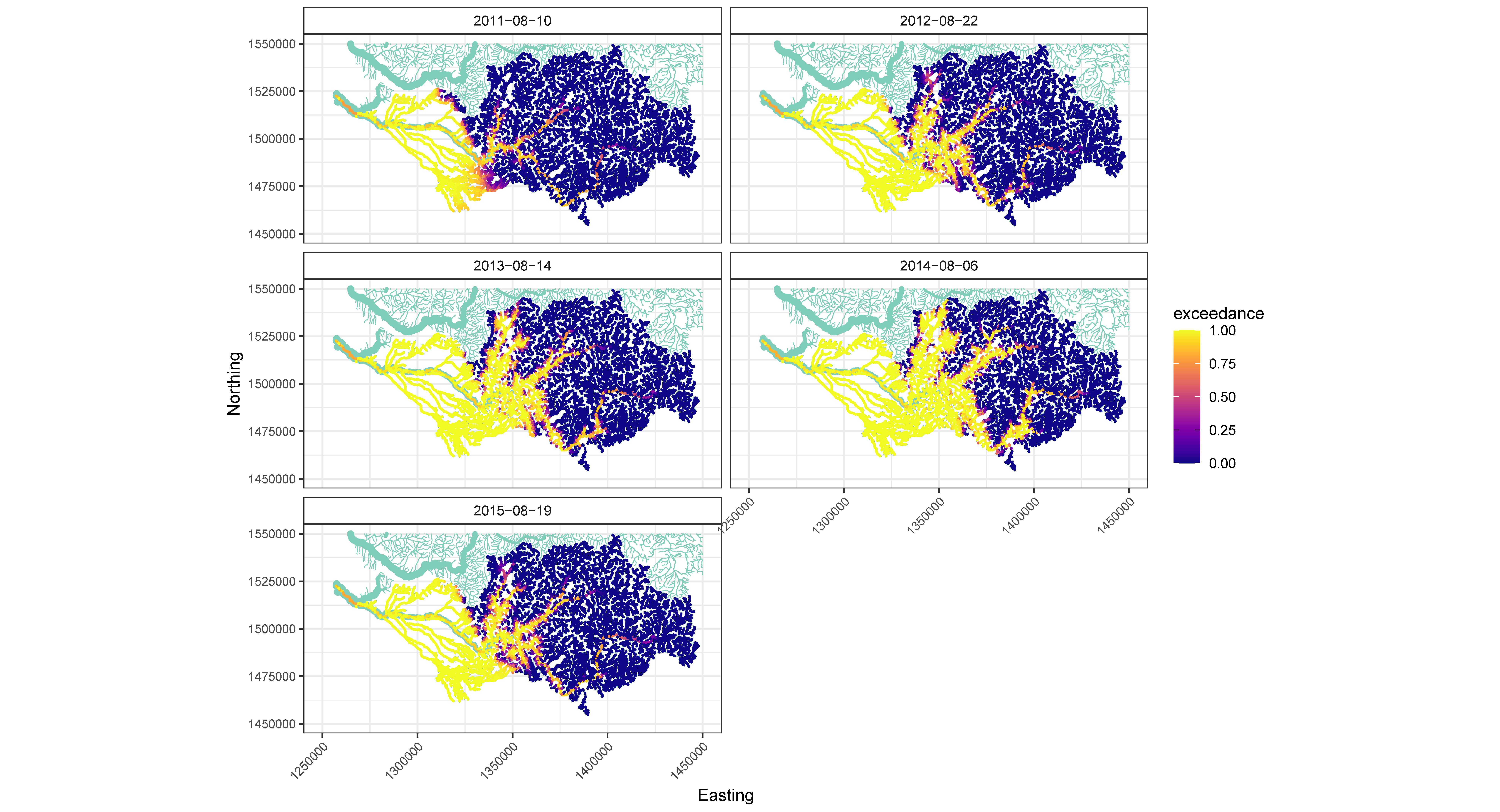}
	\caption{Probability of August mean daily temperatures exceeding 13 $^\circ$C in the Boise river network during five years of the study. } 
    \label{fig:exceedance_prob_boise}
\end{figure}

Fig \ref{fig:exceedance_prob_boise} shows the exceedance probability for days in the hottest month of the year (August) during the five years of the study. 
See an animation of the full time series in the Supplementary material. 
We note that the mainstem and lower elevation reaches of the main tributaries to the Boise River have the highest predicted probabilities.

While in Fig~\ref{fig:exceedance_prob_boise} the analysis is point by point, one attractive feature of our Bayesian analysis is that we have MCMC samples from the joint posterior predictive distribution of temperatures.  Fisheries managers and scientists are often interested in the total proportion of current and future thermal habitat unlikely to support a species (e.g. temperatures that are too high for bull trout in our example network).  The total proportion of stream length above a threshold is a nonlinear function of the joint posterior predictive distribution, but it can be estimated easily, with credibility intervals, from our MCMC sample. We computed the unsuitable proportion of habitat for bull trout based on the 13 $^\circ$C temperature limit throughout the Boise river network.
On every date, for each MCMC iteration, we calculated the proportion of the prediction points that were above the threshold.

\begin{figure}[ht]
	\centering
	\includegraphics[height=6.5in]{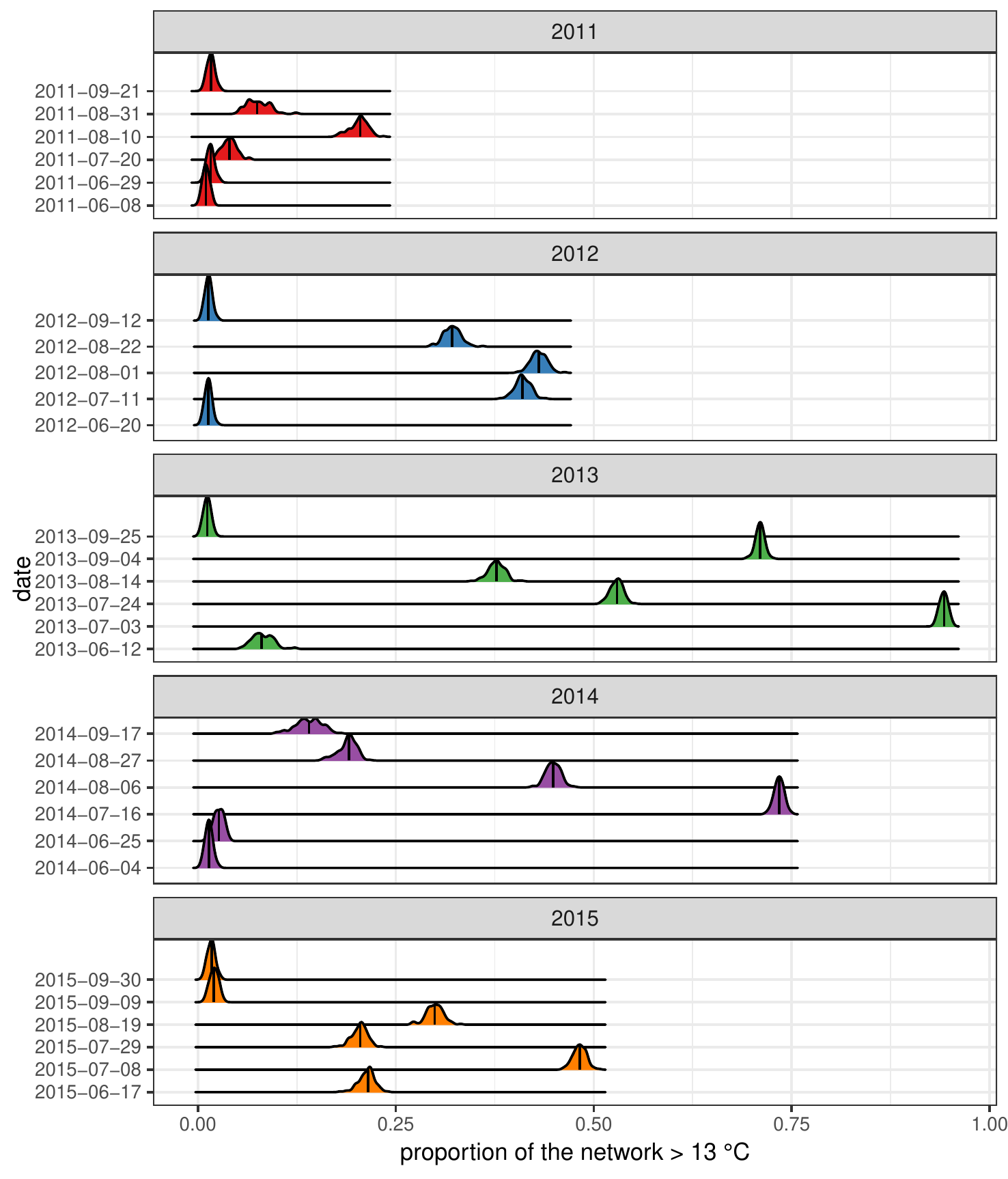}
	\caption{Posterior density of the proportion of non-suitable habitat for bull trout from June to September across the 5 years of the study. } \label{fig:prop_area}
\end{figure}

In Fig~\ref{fig:prop_area} we show the posterior densities distributions for the proportion of non-suitable habitat for bull trout during days in the hottest months of the year. The spread of these posterior densities shows the uncertainty around the estimates from the MCMC iterations.  These proportions can also be converted to total stream length impacted. The stream network was 7,364 km in total length. So, for example,  on 2013-08-14, approximately 36\% of the network was over 13 degrees, or  $\approx$ 2,651 km. 
These results are also presented in a different way (sorted by day-month) in \ref{sec:or}.

\clearpage

\section{Discussion and conclusions}
\label{sec:Dis}

Statistical modeling in stream networks is becoming an active area of research. 
This is in part, due to the rapid and widespread biodiversity losses occurring in these ecosystems. There is a critical need to assess water quality, evaluate the effects of climate change,  identify and reduce pollution levels, and understand current and future habitat suitability and species distributions.  The ability to undertake these types of analyses have only recently been possible due to the widespread use of in-situ sensors and monitoring arrays in streams, which are now generating millions of water quality measurements at thousands of sites annually. 
While we used temperature as an example here, these methods have widespread applicability for other variables collected in streams as these large datasets become more common. For example, the US Clean Water Act legally requires states and tribes to identify stream segments that exceed water-quality thresholds and to calculate their total maximum daily load based on chemical and physical water quality standards; despite the fact that it is impossible to sample all streams over time. Thus, the spatio-temporal models described here fill an important methodological gap. They can be used to generate more accurate and precise predictions for every segment within a stream, and to generate maps of exceedance probabilities, so that aquatic managers can effectively prioritize interventions and management actions in space and time.       

This paper proposes a new framework of vector autoregression spatial models for stream networks based on Bayesian inference. We have demonstrated its feasibility and have suggested several approaches and variations.
Our models are the results of extending existing spatial models to capture spatial heterogeneity in stream networks incorporating temporal dependence using vector autoregression.
We show that our approach is equivalent to separable space-time models, but it is more efficient computationally.

One of our novel model variations suggests conditioning the amount of temporal dependence on geographical parameters such as the elevation and the cumulative watershed area. 
We illustrate the use of a simple kriging method to make predictions throughout the stream network in space and time.

Estimation of the probability of exceeding certain thresholds individually or simultaneously all the spatial locations is straightforward using the MCMC outputs from the Bayesian model. This can be used to identify anomalies in water quality variables.
Imputation of missing data and interpolation is also easily done in the Bayesian context. 
All these benefits give some superiority over other methods found in the literature.

Usually, Bayesian methods are slow, especially in the context of spatial/spatio-temporal processes.  
The bottleneck is generally is having to invert large covariance matrices.
Current research is focusing on how to properly scale up similar models for big datasets in the spatial domains. See  \citet[e.g.][]{datta2016hierarchical, finley2017applying} that have suggested  Gaussian processes based on the nearest neighbours.
We are currently extending our Stan modelling framework to allow GPU computation of the likelihood, which should result in substantial efficiency gains \citep{vcevsnovar2019gpu}.

Using a site-specific autoregression parameter as in case 2a can be a limitation when making predictions using the fitted model. This issue is solved in model VAR\_2b where the autoregression parameters are regressed using available spatial covariates.
Our models can be implemented so that they are updated as new data become available \citep{stroud2001dynamic, sarkka2013bayesian, schifano2016online} and the use of previous knowledge in the form of prior distributions. 

Regarding our model comparisons in Table \ref{table:compar}, the objective was to go beyond the sole use of information criteria. 
Ultimately the decision of what model to use has to be weighted by practitioners based on their study objectives. 
The VAR\_2b approach is recommended if the aim is to produce the best possible predictions and we want to account for temporal dependence based on geographical characteristics of the site.
Practitioners seeking to fit a model that is simple and produce precise fixed effects estimates should consider AR models.

Possible extensions based on model stacking can be explored, which is a weighted average of models based for example on information criteria or uncertainty of the parameter estimates.  
Additionally, further research is required to explore models that consider the interactions between sites across different time points. Similarly, the spatial process can be defined to change smoothly over time, producing time-specific partial sills in Eq~\ref{eq:tum}-\ref{eq:edm}.
Further work needs to be done to 
develop of more efficient models and non-separable covariance matrices \citep{cressie1999classes, gneiting2002nonseparable}.

\section*{Acknowledgments}

This research was supported by the Australian Research Council (ARC) Linkage Projects: ``Revolutionising water-quality monitoring in the information age''.
This work was improved substantially based on the suggestions and recommendations from the AE and two anonymous reviewers.
We thank Dona Horan for the creation of the spatial stream network (SSN) object.
We also thank Rob Hyndman, Puwasala Gamakumara, Claire Kermorvant and members of the project from the Department of Environment and Science, Queensland for their helpful comments and feedback.

\section*{Data and Code Accessibility}

The computations were done using the \texttt{R} package \texttt{SSNbayes}, which contains the functions and \texttt{R/Stan} code for fitting the models, making predictions and visualizing stream networks. This package can be found at \url{https://github.com/EdgarSantos-Fernandez/SSNbayes}.

\bibliography{ref}


\pagebreak 
\appendix

\section{Computation of the spatial weights in the tail-up models.}
\label{sec:wei}

In this section, we illustrate how to obtain the spatial weights matrix ($W$) based on watershed area. 
Recall Fig \ref{fig:network} showing a network with four spatial locations and five stream segments. 
The coloured areas represent segment contributing areas, which is the area of land contributing overland water flow to the segment in the absence of water loss. 

\begin{figure}[htbp]
  \centering
   \includegraphics[width=4.25in]{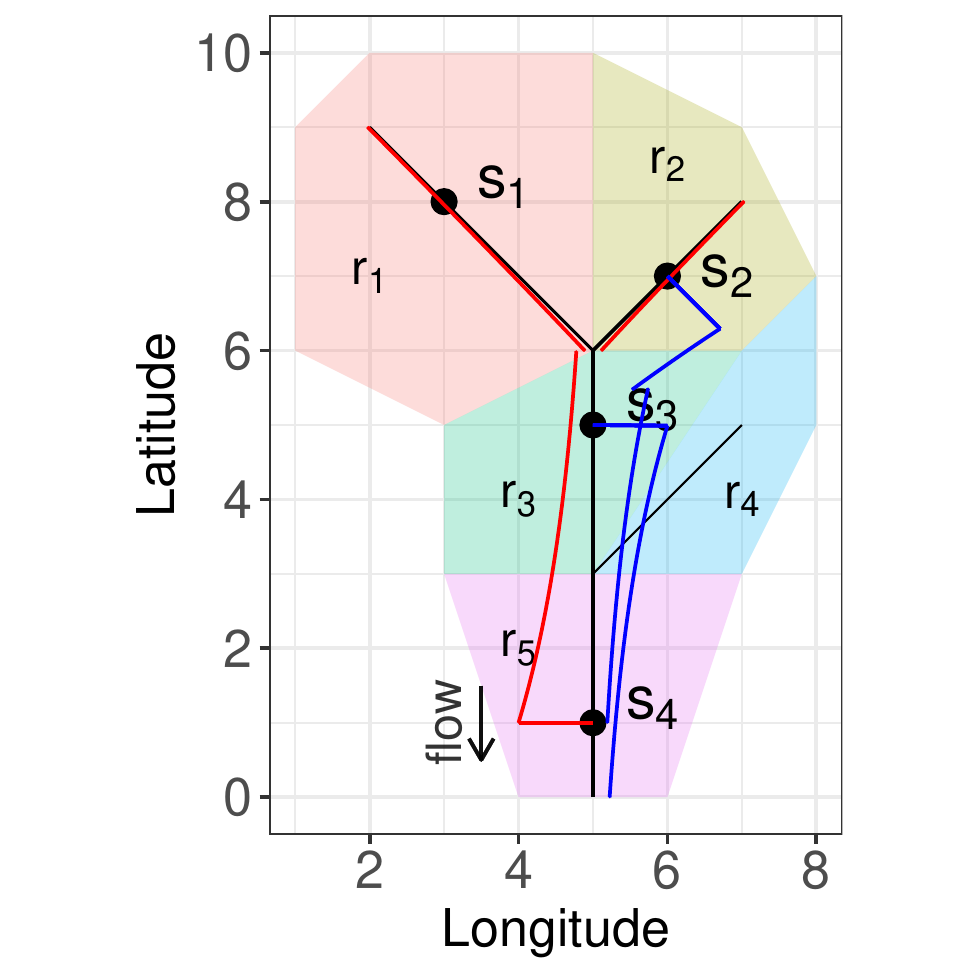}
  \caption{Stream network with four spatial locations ($s_1-s_4$) and five regions. The covariance value obtained from the moving average functions is represented with the red (tail-up) and blue (tail-down) solid segments. }
  \label{fig:networkfull}
\end{figure}

We start calculating the segment additive function value (AFV).
The first three columns in Table \ref{table:weig} are the segment id, the spatial locations and the watershed area.
The column watershed area represents the colored area around the segments in Fig \ref{fig:networkfull}.

The stream segment proportional influences (PI) are the relative influence of each segment in terms of watershed area towards a confluence. We start by assigning a PI=1 to the most downstream area. The next two areas $r_2$ and $r_3$ will have $PI_3 = 9/ (9+6.5) = 0.581$ and $PI_2 = 6.5/ (9+6.5) = 0.419$. 
The AFV is the relative importance associated to the segment in which it resides. For example $AFV_1 = PI_1 \times PI_3 \times PI_4 = 0.673 \times 0.581 \times 1 = 0.391$. Note that the AFV declines as we go further upstream in the network.

The network in Fig\ref{fig:networkfull} shows the moving average functions for tail-up and tail-down processes obtained using a partial sill equal to 0 with exponential covariance functions.
In red, from location $s_4$ we represent a tail-up model.
Notice how the function splits at the junction (lon = 5, lat=6).
The blue segments are tail-down processes going downstream from spatial locations $s_2$ and $s_3$. The height from the stream to the moving average segment represents the covariance value as a function of the distance.   

\begin{table}[ht]
\centering
\caption{Spatial weights computation}
\begin{tabular}{lrrrrr}
Segment & Location & Segment contrib areas & Watershed area & prop influence & AFV   \\\hline
r1      & s1       & 17.500                & 17.500         & 0.673          & 0.579 \\
r2      & s2       & 8.500                 & 8.500          & 0.327          & 0.281 \\
r3      & s3       & 8.000                 & 34.000         & 0.861          & 0.861 \\
r4      &          & 5.500                 & 5.500          & 0.139          & 0.139 \\
r5      & s4       & 9.000                 & 48.500         & 1.000          & 1.000
 \\\hline
\label{table:weig}
\end{tabular}
\end{table}

$\pmb{W}$ is a symmetric squared matrix indicating the weights between sites (spatial locations) and it can be obtained from an ecological spatial variable e.g. discharge. 
In Fig \ref{fig:network}, locations $s_1$ and $s_2$ are not connected by flow, then $W_{12} = W_{21} = 0$.
These weights are obtained using:

$W_{ij} =  \sqrt{\frac{AFV(s_1) }{AFV(s_2)}} $. 
For instance, the weight between spatial location
$s_1$ and $s_3$ is $w_{13} = \sqrt{\frac{AFV(s_1) }{AFV(s_3)}} =
\sqrt{\frac{0.579 }{0.861}} = 0.820$.

\[ \pmb{W} = 
\begin{blockarray}{ccccc}
   & s1    & s2    & s3    & s4    \\
\begin{block}{c(cccc)}
s1 & 1     & 0     & 0.820 & 0.761 \\
s2 & 0     & 1     & 0.572 &0.530 \\
s3 & 0.820 & 0.572 & 1     & 0.928 \\
s4 & 0.761  & 0.530  &0.928  & 1  \\
\end{block}
\end{blockarray}
 \]

\section{Hierarchical model and prior distributions.} 
\label{sec:hierar}


{\scriptsize \centering
\begin{align*}
[\pmb{y}_1,\pmb{y}_2,\cdots,\pmb{y}_T] &=  \prod_{t=2}^{T}[\pmb{y}_t \mid  \pmb{y}_{t-1} ,\pmb{\theta},\pmb{X}_{t},\pmb{X}_{t-1},\pmb{\beta}, \pmb{\Phi}_1, \pmb{\Sigma}][\pmb{y}_1]  \\
[\pmb{y}_t \mid  \pmb{y}_{t-1} ,\pmb{\theta},\pmb{X}_{t},\pmb{X}_{t-1},\pmb{\beta}, \pmb{\Phi}_1, \pmb{\Sigma}] &= \mathcal{N}(\pmb{\mu}_{t},\pmb{\Sigma} + \sigma_{0}^{2} \pmb{I} ) \\
\pmb{\mu}_{t} &= \pmb{X}_{t}\pmb{\beta} + \pmb{\Phi}_1 (\pmb{y}_{t-1} - \pmb{X}_{t-1}\pmb{\beta}) \\
\pmb{\Sigma} & = \sigma^2_{u}\pmb{R}(\alpha_u) + \sigma^2_{d}\pmb{R}(\alpha_d)+\sigma^2_{e}\pmb{R}(\alpha_e)  && \\
\textrm{Priors}  && \\
    \beta_0, \beta_1,\beta_2, \cdots, \beta_p
     & \sim \mathcal{N}\left(0,100\right)  && \text{\# prior on the regression coefficients}\\	
    \sigma_{0} & \sim \textrm{Uniform}(0,50) && \text{\# prior on the nugget effect}\\
	\sigma_{u}, \sigma_{d}, \sigma_{e} & \sim \textrm{Uniform}(0,100) && \text{\# prior  partial sill parameters }\\
	\alpha_{u}, \alpha_{d}, \alpha_{e}  & \sim \textrm{Uniform}(0,\alpha_{max}) && \text{\# prior on spatial range parameters }\\
	\alpha_{max} &= 4 \max(H) && \text{\# Four times the maximum stream distance. }   \\
	\textrm{Elements of }  \pmb{\Phi}_1 \textrm{:} &&  \\
    \textrm{Case 1}  &&  \\
	\phi &\sim \textrm{Uniform}(-1,1) &&  \text{\# prior on the 
	autoregressive parameters}\\	
	\textrm{Case 2a}  &&  \\
	\phi_{s} &\sim \mathcal{N}(0.5, 0.2)T[-1,1] &&  \text{\# Truncated norma prior on the site specific autoregressive parameters}\\	
	\textrm{Case 2b}  &&  \\
	\textrm{logit}(\phi_s) &= \gamma_0 + \gamma_1  X_{1s} + \gamma_2  X_{2s} + \cdots + \gamma_J  X_{JS}  &&  \text{\# prior on the site specific autoregressive parameters}\\
	\gamma_0, \gamma_1,\gamma_2, \cdots, \gamma_J &\sim \mathcal{N}(0, 100) &&  \text{\# regression coefficients for autoregressive}\\
	& &&  \text{  parameters for location } s = 1,2,\cdots,S.\\	
	\textrm{Case 3}  &&  \\
	\phi_{s}, \phi_{sr} &\sim \textrm{Uniform}(-1,1) &&  \text{\# prior on the autoregressive parameters for  location} s = 1,2,\cdots,S \text{ and }\\	
	 & &&  \text{  its 2 NN neighbours} r = 1,2,\cdots,R.\\	
\end{align*} 
}
\label{fig:priors}

The upper limit for the spatial range prior ($\alpha_{max}$) is computed as four times the longest distance between spatial locations.

\section{Other results} 
\label{sec:or}

\begin{figure}[ht]
	\centering
	\includegraphics[width=7.25in]{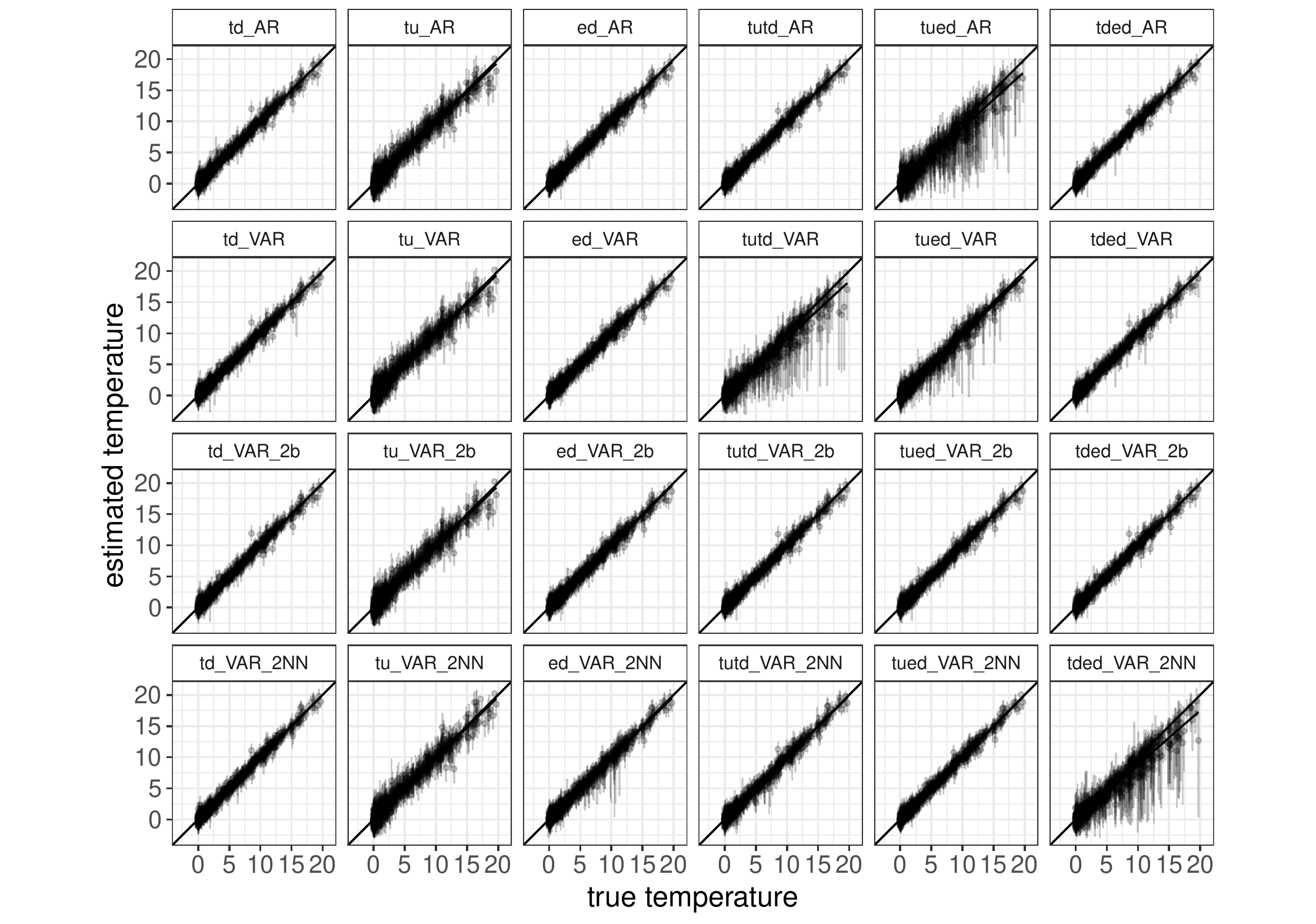}
	\caption{Comparison of the estimated (posterior mean) versus the true latent stream temperature values from the 24 models for the 87 dates used in the analysis. The vertical lines are the 95\% highest density intervals (HDI).  } 
    \label{fig:temp_pred_vs_obs_HDI}
\end{figure}

\begin{figure}[ht]
	\centering
	\includegraphics[width=3.5in]{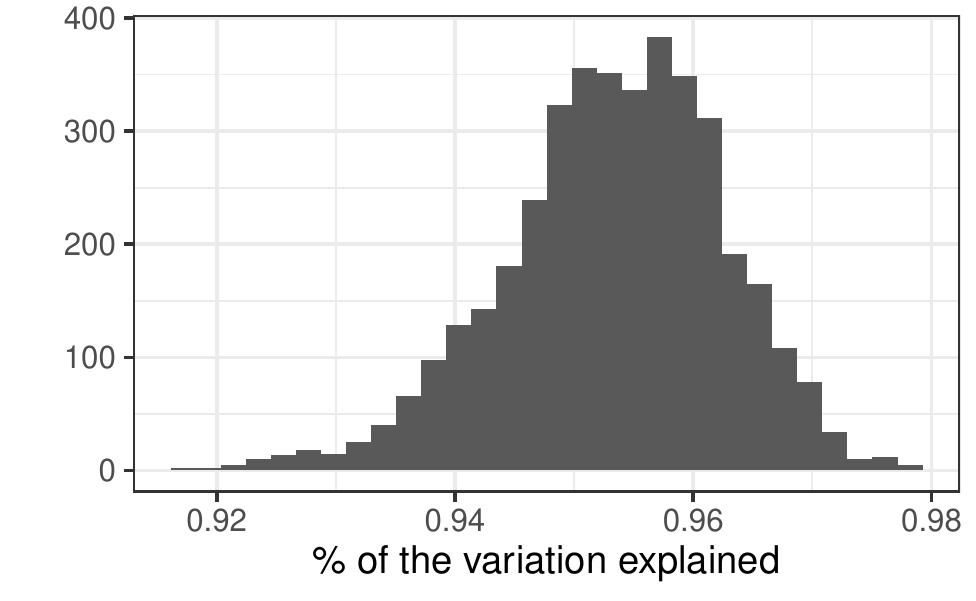}
	\caption{Percentage of the variation explained ($\sigma_{td} / (\sigma_{td} + \sigma_{0}$)) from the td\_VAR\_2b model  } 
    \label{fig:perc_var_explained}
\end{figure}

\begin{figure}[ht]
	\centering
	\includegraphics[height=3.5in]{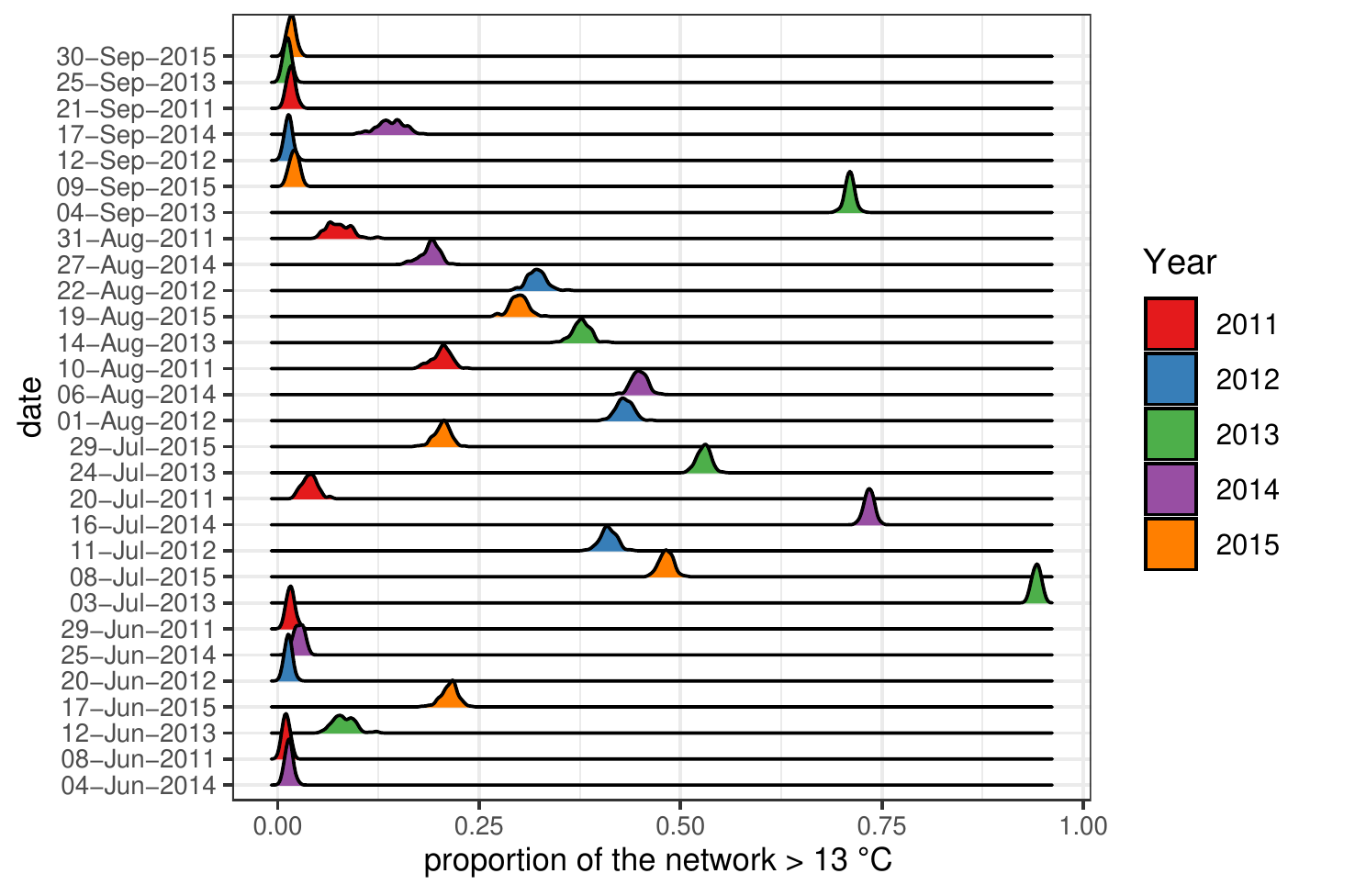}
	\caption{Posterior density of the proportion of non-suitable habitat for bull trout from June to September across the 5 years of the study. The distributions are sorted by day-month and the years are represented in different colors.} \label{fig:prop_area_sorted}
\end{figure}

\clearpage

\section{Variance, Covariance, and Correlation for the VAR(1) common $\phi_s$} 
\label{sec:invers}

In this section, we develop the variance, covariance, and correlation in the VAR(1) with common $\phi$ values (Case 1).



Let $\by_{t}$ be a vector of random variable at time $t$, and consider the stochastic process $\{\by_{t}; t = 0, 1, \ldots, T \}$. Consider the stochastic VAR(1) process, 
$$
\by_{t} = \phi \by_{t-1} + \bepsilon_{t},
$$
where $\phi$ is a scalar and $\bepsilon_{t}$ is independent from any $\bepsilon_{u}$ where $u \ne t$. Without loss of generality, we assume that $\mathbb{E}(\bepsilon_{t}) = \bzero  \ \forall \ t$ (as a mean can always be added later) and $\var(\bepsilon_{t}) = \bV$. It is easy to see that the stochastic process $\{\by_{t}\}$ is stationary in the mean, $\mathbb{E}(\by_{t}) = 0  \ \forall \ t$. We want the variance to be stationary as well, $\var(\by_{t}) = \var(\by_{t+h}) \ \forall \ t,h$. The variance of $\by_{t}$ is
$$
\var({\by_{t}}) = \phi^{2} \var(\by_{t-1}) + \bV,
$$
and we want the conditions where $\var({\by_{t}}) = \var(\by_{t-1})$, which occurs at 
$$
\var({\by_{t}}) = \frac{1}{1 - \phi^{2}}\bV.
$$
Note that this requires $\phi^{2} < 1$ and setting $\var(\by_{0}) = (1 - \phi^{2})^{-1}\bV$.

For the covariance of ${\by_{t}}$ and $\by_{t+h}$, note that, by substitution
$$
\by_{t+2} = \phi(\phi \by_{t} + \bepsilon_{t+1}) + \bepsilon_{t+2} = \phi^{2} \by_{t} + \phi \bepsilon_{t+1} + \bepsilon_{t+2},
$$
and continuing with substitution in this fashion, we can generalize for any $h$,
$$
y_{t+h} = \phi^{h}\by_{t} + \sum_{i=0}^{h-1}\phi^{i}\bepsilon_{t+h-i}.
$$
Then 
$$
\cov(\by_{t},\by_{t+h}) = \cov(\by_{t},\phi^{h}\by_{t} + \sum_{i=0}^{h-1}\phi^{i}\bepsilon_{t+h-i}),
$$
but $\cov(\by_{t}, \bepsilon_{t+h-i}) = 0$ for all $i$ as all $\bepsilon_{t+h-i}$ are in the future of $\by_{t}$.  Hence,
$$
\cov(\by_{t},\by_{t+h}) = \cov(\by_{t}, \phi^{h}\by_{t}) = \phi^{h}\var(\by_{t}) = \frac{\phi^{h}}{1 - \phi^{2}}\bV.
$$
It is also obvious that 
$$
\corr(\by_{t},\by_{t+h}) = \phi^{h}\bI.
$$

\newpage
\noindent \textbf{Inverse of Covariance Matrix for VAR(1) Model -- Scalar $\phi$}
\vspace{.5cm}

\vspace{.3cm}

\vspace{.3cm}
We can write the covariance matrix using Kronecker products as

\begin{equation} \label{eq:VAR1covmat_phiI}
\var(\by) \equiv \bSigma_{var1} = \frac{1}{1 - \phi^{2}}
\left(
\begin{array}{ccccc}
  \bV & \phi\bV & \phi^{2}\bV & \cdots & \phi^{T}\bV \\
  \phi\bV & \bV & \phi\bV & \cdots & \phi^{T-1}\bV \\
  \phi^{2}\bV & \phi\bV & \bV & \cdots & \phi^{T-2}\bV \\
  \vdots & \vdots & \vdots & \ddots & \vdots \\
  \phi^{T}\bV & \phi^{T-1}\bV & \phi^{T-2}\bV & \cdots & \bV
\end{array}
\right) = \bSigma_{ar1} \otimes \bV
\end{equation}
where $\bSigma_{ar1}$ is the same as Equation~\eqref{eq:AR1covmat} where $\sigma^{2} = 1$ (which has essentially been replaced by $\bV$).  Recall that $(\bA \otimes \bB)^{-1} = \bA^{-1} \otimes \bB^{-1}$, so,
\begin{equation} \label{eq:VAR1covInverse}
\bSigma_{var1}^{-1} = \bSigma_{ar1}^{-1} \otimes \bV^{-1} = 
\left(
\begin{array}{cccccc}
  \bV^{-1} & -\phi\bV^{-1} & \bzero & \cdots & \bzero & \bzero \\
  -\phi\bV^{-1} & (1 + \phi^{2})\bV^{-1} & -\phi\bV^{-1} & \cdots & \bzero & \bzero \\
  \bzero & -\phi\bV^{-1} & (1 + \phi^{2})\bV^{-1} & \cdots & \bzero & \bzero \\
  \vdots & \vdots & \vdots & \ddots & \vdots & \vdots \\
  \bzero & \bzero & \bzero & \cdots & (1 + \phi^{2})\bV^{-1} & -\phi\bV^{-1} \\
  \bzero & \bzero & \bzero & \cdots & -\phi\bV^{-1} & \bV^{-1}
\end{array}
\right).
\end{equation}

\section{Simulation study comparing both spatio-temporal methods and their predictions.}
\label{sec:sim}

In this section, we compare the full separable space-time covariance model and the vector autoregression spatial approach using simulated data on the usual spatial settings (no in stream networks).

{\bf Method 1: Separable space-time VAR(1) model }

Space-time autocorrelation is incorporated in the separable covariance matrix. It takes advantage of the properties of  the Kronecker product. Let

$$
[\pmb{y}_t \mid  \pmb{y}_{t-1} ,\pmb{\theta},\pmb{X}_{t},\pmb{X}_{t-1},\pmb{\beta}, \pmb{\Phi}_1, \pmb{\Sigma}] = \mathcal{N}(\pmb{\mu}_{t},\pmb{\Sigma} + \sigma_{0}^{2} \pmb{I} ),$$

$$
\pmb{\mu}_{t} = \pmb{X}_{t}\pmb{\beta} 
\label{eq:errmu},
$$

\noindent The inverse of the separable covariance matrix is 

$$
\pmb{\Sigma}^{-1} = \pmb{\Sigma}_{S}^{-1} \otimes \pmb{\Sigma}_{var}^{-1}
$$

$$$$
\noindent where $\pmb{\Sigma}_{S}$ is the spatial covariance matrix defined in Eq~\ref{eq:covs} and $\pmb{\Sigma}_{var}$ is the temporal covariance matrix of the VAR(1) process.

This property reduces substantially the computation time since we just need to invert the covariance matrices of space and time and then multiply them.

{\bf  Method 2: Vector autoregression spatial model}

In this approach we just need to construct the spatial covariance matrix. 
Temporal dependence is incorporated in the error terms using a vector autoregression VAR(1).
Let,

\begin{equation}
[\pmb{y}_t \mid  \pmb{y}_{t-1} ,\pmb{\theta},\pmb{X}_{t},\pmb{X}_{t-1},\pmb{\beta}, \pmb{\Phi}_1, \pmb{\Sigma}] = \mathcal{N}(\pmb{\mu}_{t},\pmb{\Sigma} + \sigma_{0}^{2} \pmb{I} ),
\end{equation}

\begin{equation}
\pmb{\mu}_{t} = \pmb{X}_{t}\pmb{\beta} + \pmb{\Phi}_1 (\pmb{y}_{t-1} - \pmb{X}_{t-1}\pmb{\beta}). 
\label{eq:errmu}
\end{equation}

The objectives of this study are to

\begin{enumerate}
    \item show that both approaches are equivalent
    \item compare them in terms of computational efficiency
    \item explore the options for producing prediction (kriging) and assess the model interpolation/imputation. 
\end{enumerate}

\begin{figure}[ht]
	\centering
	\includegraphics[width=3.25in]{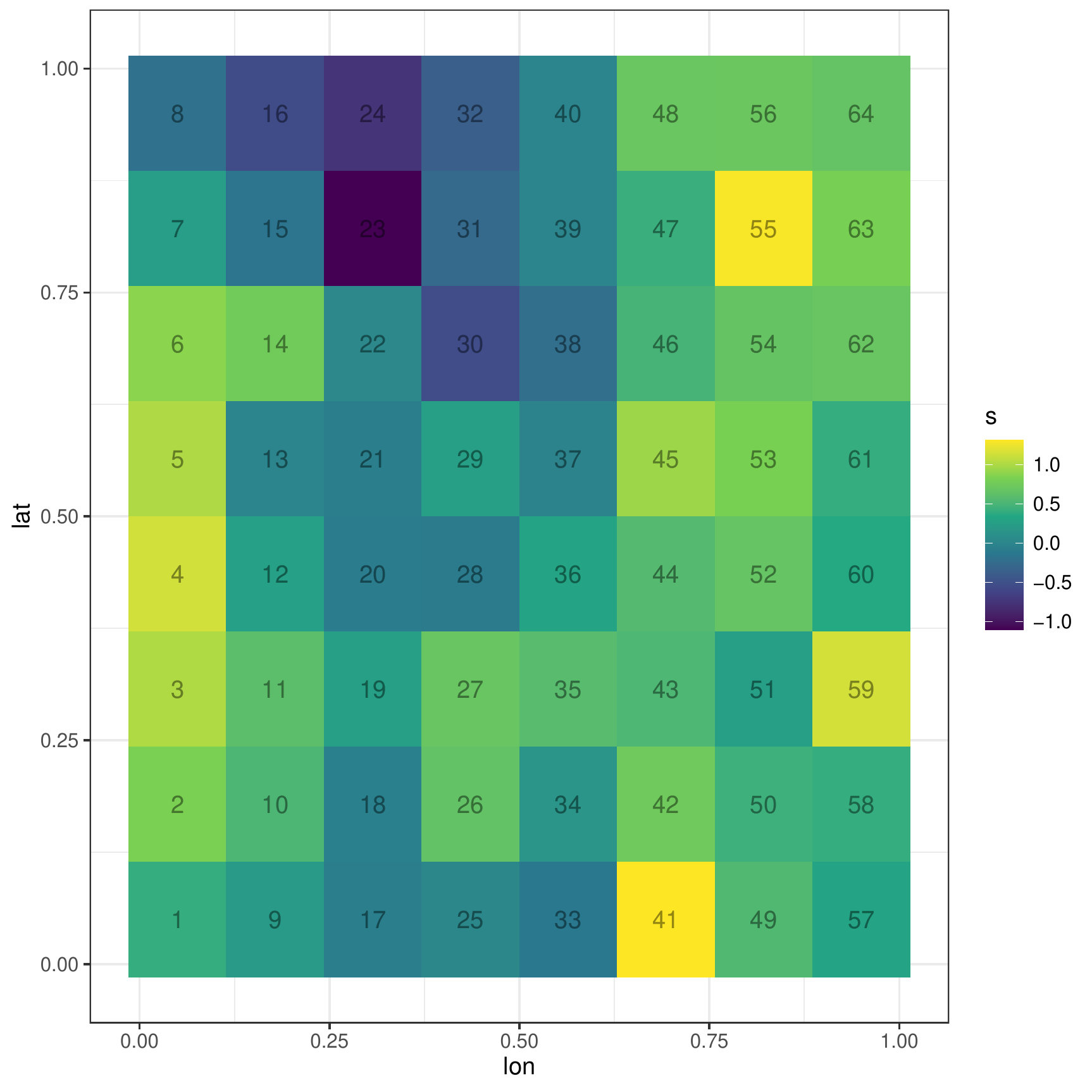}
	\caption{Spatial surface } 
    \label{fig:ss}
\end{figure}

We simulated a surface composed by $s = 64$ spatial locations using an exponential covariance matrix with $\sigma_{ed} = 1$ and $\alpha_{ed} = 1$. See Fig \ref{fig:ss}, where the label is the spatial location id.
Let us consider one covariate $x$ and the let us fix the regression coefficients $\beta_0 = -1$ and $\beta_1 = 2$. 

\begin{figure}[ht]
	\centering
	\includegraphics[width=3.25in]{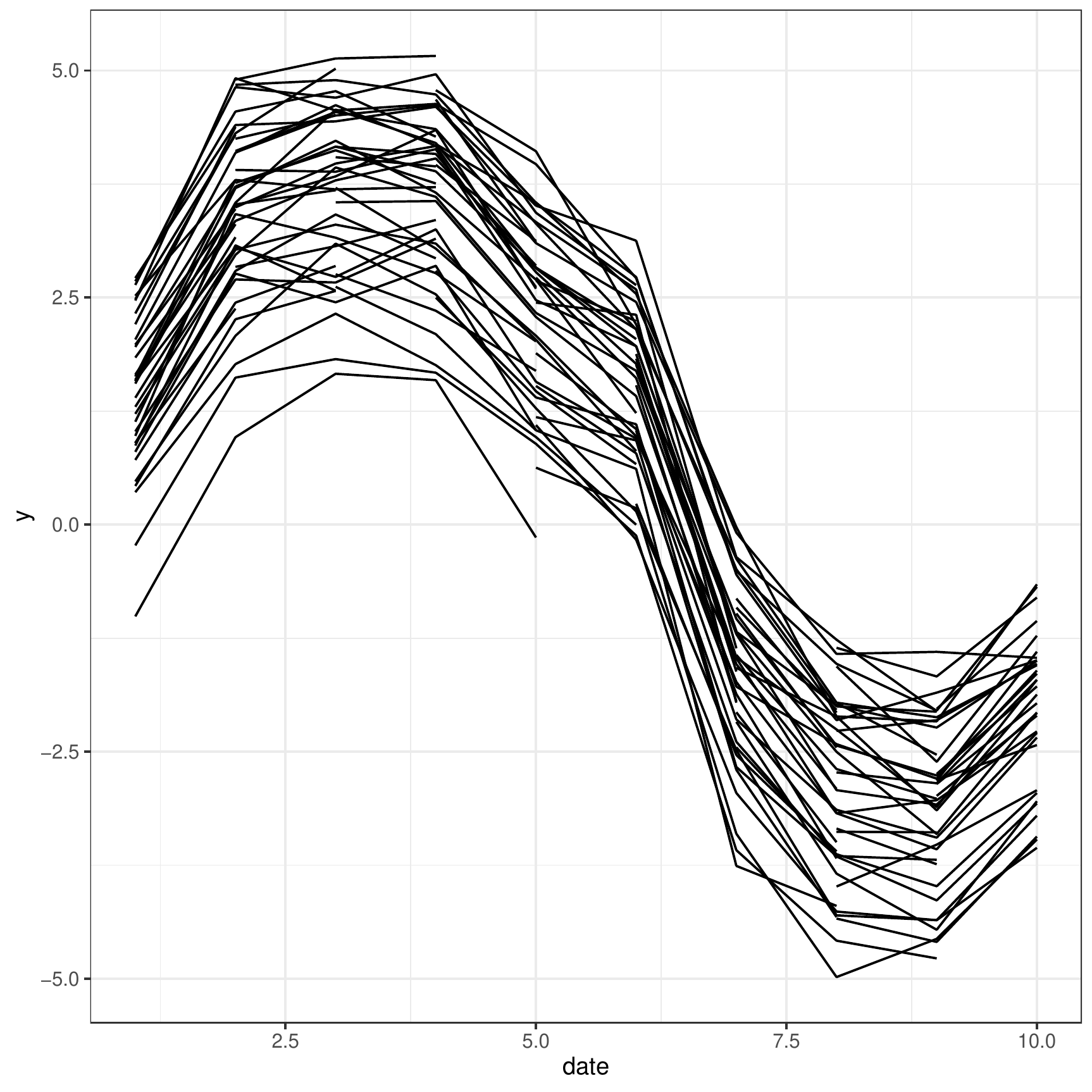}
	\caption{Time series } 
    \label{fig:ts}
\end{figure}

We generated time series using $t = 10$ time points and an AR(1) covariance matrix using $\phi = 0.6$ to obtain the error term $\pmb{\epsilon}$. 
We obtained the response variable $\pmb{y} = \pmb{X}\pmb{\beta} + \pmb{v} + \pmb{\epsilon}$, where $\pmb{v}$ is the spatial component, and then added the error $\pmb{\epsilon}$.
The resulting time series are shown in Fig~\ref{fig:ts}.

\begin{figure}[ht]
	\centering
	\includegraphics[width=7.8in]{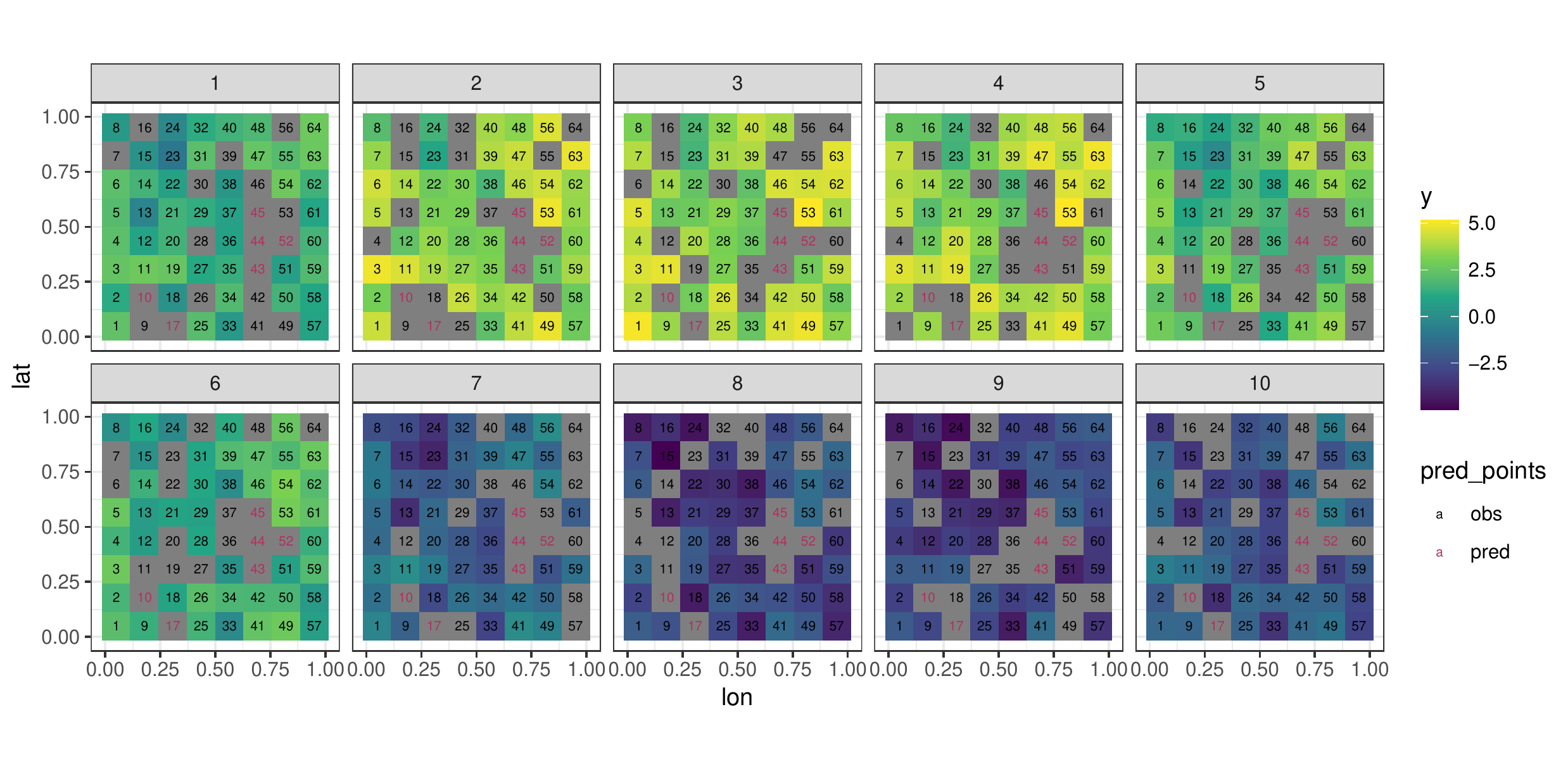}
	\caption{Spatio-temporal dataset.
	Gray tiles represent missing values (testing set) and the red labels represent the six spatial locations that will be used for kriging/prediction.} 
    \label{fig:spat_temp_sim}
\end{figure}

We created a testing set by setting to missing the full time series in 6 spatial locations (6 locations $\times$ 10 days = 60 predictions points). Here we will assess the prediction accuracy of the model.
Additionally, in the other $64-6 = 58$ locations we randomly selected observations and set them to missing to assess the quality of the imputation/interpolation.
Fig \ref{fig:spat_temp_sim} shows the evolution of the response variable across the 10 time points. In gray we give the missing values and the red labels represent the six spatial locations that will be used for kriging/prediction.

We fitted the full covariance matrix and the vector autoregression spatial model using Stan. 
In the first model we obtained the written the temporal AR(1) covariance matrix as,
\begin{equation} \label{eq:AR1covmat}
\pmb{\Sigma}_{var} = \frac{\sigma^{2}}{1 - \phi^{2}}
\left(
\begin{array}{ccccc}
  1 & \phi & \phi^{2} & \cdots & \phi^{T} \\
  \phi & 1 & \phi & \cdots & \phi^{T-1} \\
  \phi^{2} & \phi & 1 & \cdots & \phi^{T-2} \\
  \vdots & \vdots & \vdots & \ddots & \vdots \\
  \phi^{T} & \phi^{T-1} & \phi^{T-2} & \cdots & 1
\end{array}
\right) ,
\end{equation}
\noindent where $T$ is the number of time points.
In both models, we used an exponential spatial covariance matrix based on Euclidean distance.

\begin{equation}
    C_{ED} = \sigma_{e}^2 e^{-3d/\alpha_{e}}
    \label{eq:ced__}.
\end{equation}

Both Bayesian models produce imputation of the missing values in the testing set including the prediction in the six locations that the time series was set to missing. 

\begin{figure}[ht]
	\centering
	\includegraphics[width=5.5in]{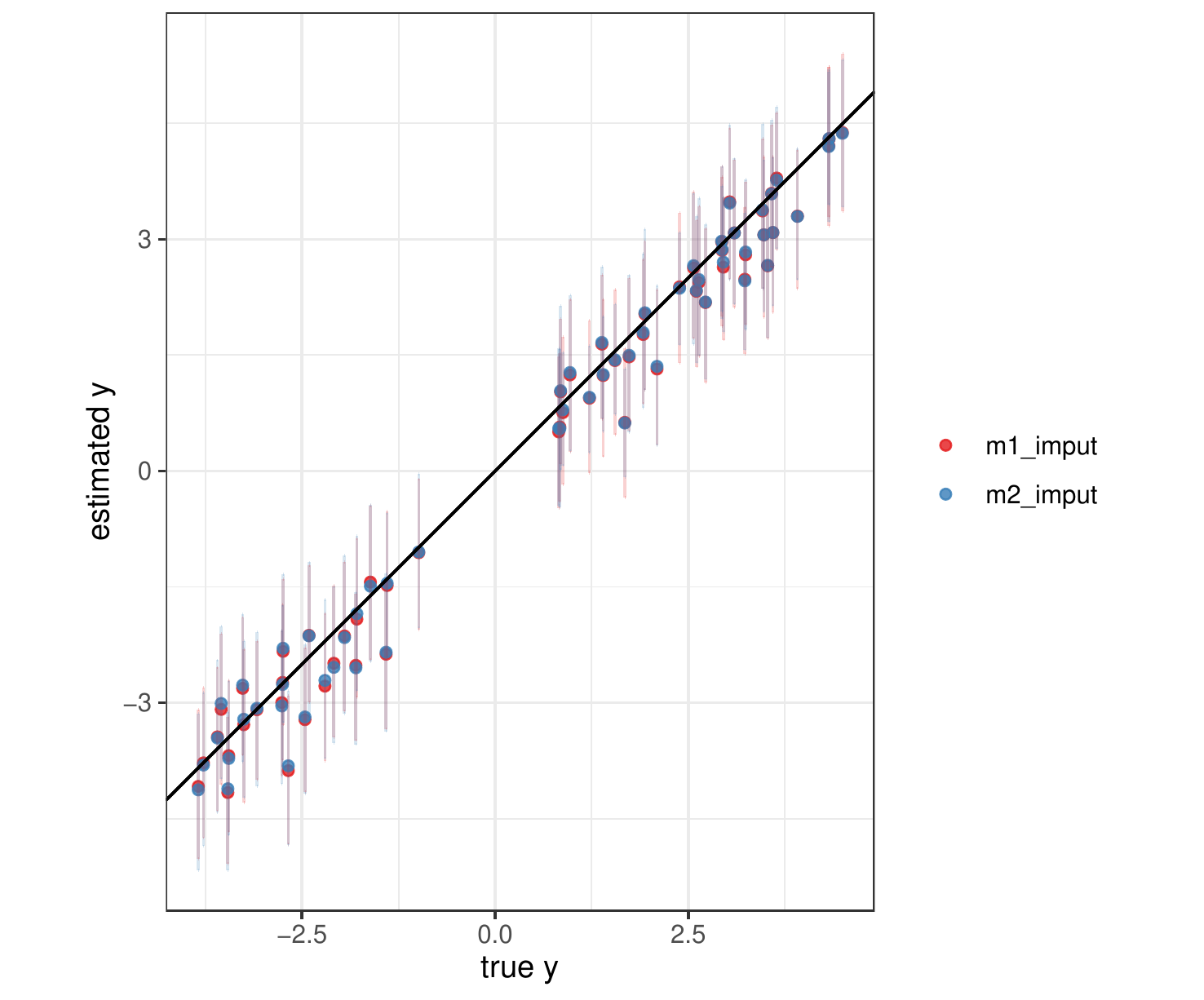}
	\caption{True vs estimated response variable. The full method is green while the vector autoregression spatial approach in red.  
	The vertical bars give the 95\% highest density intervals.} 
    \label{fig:4true_vs_pred}
\end{figure}

Fig \ref{fig:4true_vs_pred} depicts the true latent vs estimated response variable in the testing set (out-of-sample prediction). The full method is green while the vector autoregression spatial approach in red. 
As expected both methods produce very similar estimates and uncertainty. 

However, the computing time in the full method was 1.26 hrs vs 0.16 hr (10 mins) in the vector autoregression spatial one, despite the use of the Kronecker properties.

\subsection{Predictions}

Often in spatial modelling is desired to make predictions on several thousand locations. This results in a high dimensional covariance matrix that has to be inverted at every MCMC sweep, making the model prohibitive. 
A simple solution for this is to use a two stage approach: (1) fit the model to the observed data and (2) produce predictions using a simple kriging method to subsets of the data.
Predictions with the different subsets can be obtained in parallel or in different machines without communication. 
This process is relatively fast since it only involves inverting the covariance matrix between observed locations. 

To make the predictions we use the posterior distributions of the parameters of interest. We take a random sample of the stacked chains of the parameters e.g. 1000 samples. 

For the simulation study, we produced predictions in the six spatial location in Fig \ref{fig:spat_temp_sim} in red color using both methods.

\vspace{1cm}
{\it Predictions of method 1 }

In this approach we need to construct the full separable space-time covariance matrix between the observed points $C_{OO}$ and the covariance between prediction and observed points $c_{OP}$. The bottleneck is generally computing $C^{-1}_{OO}$, which in practice is not too expensive.  

\begin{equation}
\widehat{\pmb{y}}_{P} = \pmb{\mu}_{P} + {c_{OP}}'C^{-1}_{OO} (\pmb{y}_{O} - \pmb{\mu}_{O} )
\label{eq:gen}
\end{equation}

\noindent where $\pmb{\mu}_{P} = \pmb{X}_{P}\pmb{\beta}$ , $\pmb{\mu}_{O} = \pmb{X}_{O}\pmb{\beta}$, and $O$ and $P$ denotes observation and prediction location respectivelly. 

\vspace{1cm}

{\it Predictions of method 2}

In the vector autoregression spatial method we add to the observed and predicted means ($\pmb{\mu}$) in Eq \ref{eq:gen} the residual times $\pmb{\Phi}$ :   

\begin{equation}
    \pmb{\mu}_{P_{t}} = {\pmb{X}_{P_{t}}}\pmb{\beta} + \pmb{\Phi}(\pmb{y}_{P_{t-1}} - \pmb{X}_{P_{t-1}}\pmb{\beta}) 
\end{equation}

\begin{equation}
    \pmb{\mu}_{O_{t}} = \pmb{X}_{O_{t}}\pmb{\beta} + \pmb{\Phi}(\pmb{y}_{O_{t-1}} - \pmb{X}_{O_{t-1}}\pmb{\beta})
\end{equation}

\begin{figure}[ht]
	\centering
	\includegraphics[width=6.5in]{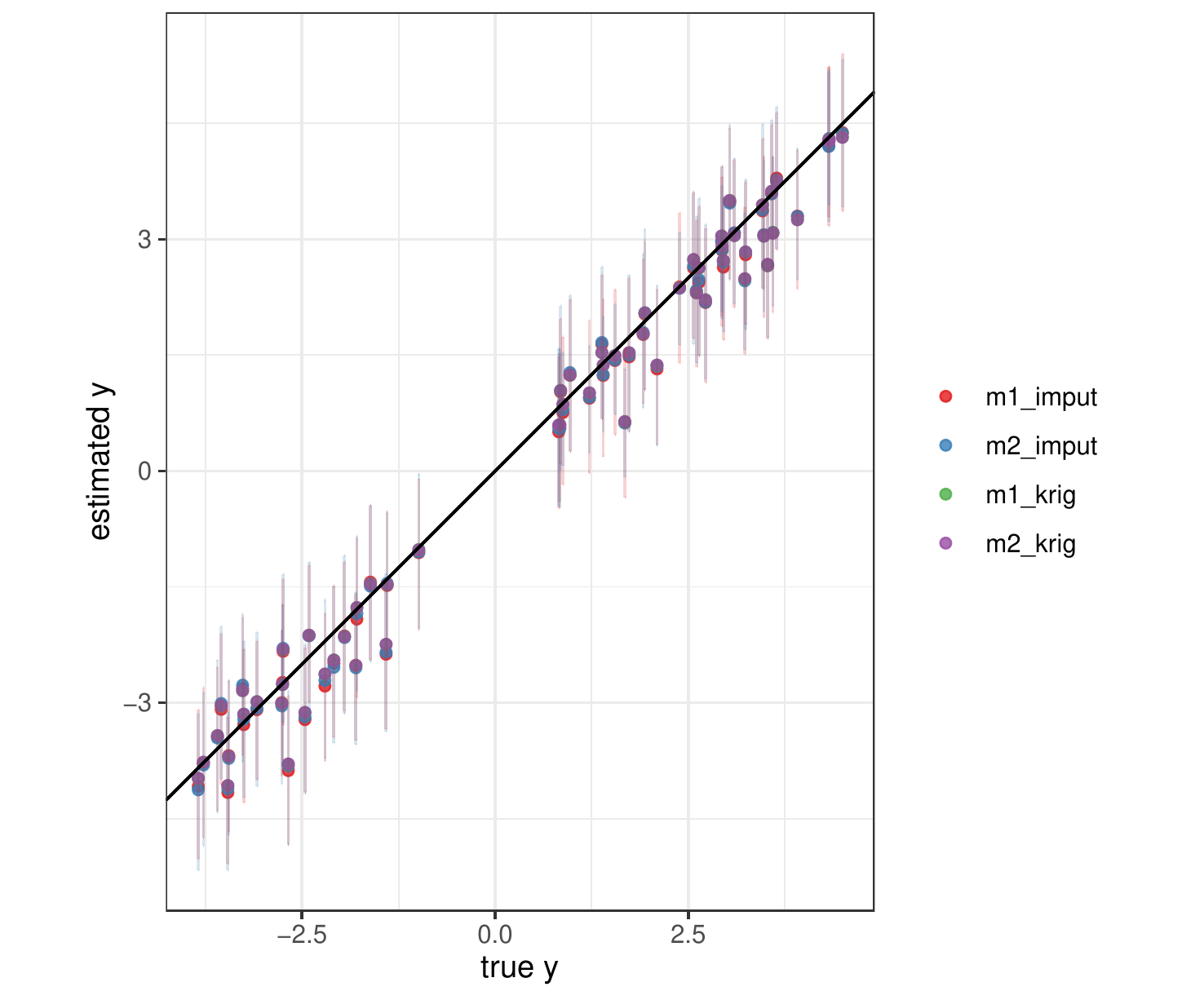}
	\caption{True vs estimated response variable. The full method is red while the vector autoregression spatial approach in blue.
	The predictions using the simple kriging from both methods are also shown in green and purple.  
	The vertical bars give the 95\% highest density intervals.} 
    \label{fig:5true_vs_pred_krig}
\end{figure}

Fig \ref{fig:5true_vs_pred_krig} shows the predicted response variable using the kriging methods and the imputation approaches previously shown in  
Fig \ref{fig:4true_vs_pred}. 
The RMSPE in both methods and the kriging predictions are similar (Table \ref{table:rmspefin}). 
Computationally, there was not much difference between the prediction methods for this small dataset.

\begin{table}[ht]
\centering
\caption{Root Mean Square Prediction Error (RMSPE).}
\begin{tabular}{lr}
method & RMSPE  \\ \hline
imputation method 1     & 0.418 \\
imputation method 2     & 0.424 \\
prediction using kriging on method 1     & 0.399 \\
prediction using kriging on method 2     & 0.399 \\ \hline
\label{table:rmspefin}
\end{tabular}
\end{table}

Summing up, this simulation study demonstrates that both methods are similar in terms of RMSPE, but the vector autoregression spatial approach is computationally more efficient.
Also, that making predictions or imputing the values using either method 
yield the same results. 
If the number of prediction locations is small compared to the observation locations imputing the values would be preferable, however, this is rarely the case.
In the presence of a large number of prediction locations, a two-stage approach is preferable: fitting the vector autoregression spatial model to the observed data and producing predictions by parts. 
\end{document}